\documentclass[a4paper,10.9pt]{article}
\usepackage[margin=3.5cm]{geometry}
\usepackage{graphicx}
\usepackage{float}
\usepackage{url}
\usepackage{amsmath}
\usepackage{gensymb}
\usepackage[utf8]{inputenc}
\usepackage{enumitem}
\usepackage{tikz}
\usetikzlibrary{decorations.pathmorphing,backgrounds,positioning,fit,arrows}
\usepackage{calc}
\usepackage{pgfplots}
\usepackage{caption}
\usepackage{subcaption}
\usepackage{xparse}
\usepackage{physics}
\usepackage{amssymb}
\usepackage{multicol}
\usepackage{color}
\usepackage{comment}
\usepackage{pgfplots}
\usepackage{titlesec}
\usepackage[square,sort&compress,numbers]{natbib}
\numberwithin{equation}{section}

\titleformat*{\section}{\large\bfseries}
\titleformat*{\subsection}{\itshape}

\DeclareMathAlphabet\mathbit
    \encodingdefault\rmdefault\bfdefault\itdefault
\DeclareOldFontCommand{\bi}{\normalfont\bfseries\itshape}{\mathbit}

\newcommand{\I}{\mathrm{i}}
\renewcommand{\vec}[1]{{\boldsymbol{#1}}}
\DeclareMathOperator{\diag}{diag}
\pgfplotsset{compat=1.14}

\title{\hfill\\
\LARGE\textbf{Dynamic Green's functions in discrete\\ flexural systems}}
\author{\hfill\\
K.~H.~MADINE \scshape{and} D.~J.~COLQUITT\\
\small \textit{ Department of Mathematical Sciences, University of Liverpool,} L69 7ZL, UK\\
\hfill}

\date{}

\begin{document}

\maketitle

\begin{abstract}
\noindent\small The paper presents an analysis of the dynamic behaviour of discrete flexural systems composed of Euler--Bernoulli beams.
The canonical object of study is the discrete Green's function, from which information regarding the dynamic response of the lattice under point loading by forces and moments can be obtained.
Special attention is devoted to the interaction between flexural and torsional waves in a square lattice of Euler--Bernoulli beams, which is shown to yield a range of novel effects, including extreme dynamic anisotropy, non-reciprocity, wave-guiding, filtering, and the ability to create localised defect modes, all without the need for additional resonant elements or interfaces.
The analytical study is complimented by numerical computations and finite element simulations, both of which are used to illustrate the effects predicted.
A general algorithm is provided for constructing Green's functions as well as defect modes.
This algorithm allows the tuning of the lattice to produce pass bands, band gaps, resonant modes, wave-guides, and defect modes, over any desired frequency range.
\end{abstract}
\hfill


\section{Introduction}\label{Intro}
In recent years, the study of lattice dynamics has undergone something of a renaissance due, in part, to the counter-intuitive effects associated with metamaterials, such as perfect lenses~\cite{Pendry2000}, negative refractive index materials~\cite{Shelby2001}, and invisibility cloaks~\cite{Pendry2006}.
The majority of scholarly work in this area has focused on photonic, plasmonic, and acoustic metamaterials, whilst the science of mechanical metamaterials attracts less attention.
Nevertheless, elastic metamaterials have found applications a wide array of settings, from cloaking to energy dissipation in engineering structures and seismic protection, among many others~\cite{CymanticsCloaking,TransformElastoCloak,AchieveControlCloak,SeismicMetawedge,CoupledStressCloak,CartaPhononBGofBridge,GyroVibrationReduction,MetamatOverview}.

Much of the work on metamaterials is underpinned by the study of wave propagation in periodic media, which has been studied in various forms for centuries and remains an active area of research today~\cite{Brillouin,lattice-sums,slepyan2012models}.
Dynamic vector systems corresponding to multi-scale mechanical materials offer unique opportunities to study the flexural and rotational displacements that occur in elastic lattice systems~\cite{MartinssonVibratePhononic,SlepyanCrack,ParabolicMetamaterials,ChiralWaves,DynamicAnisotropy,RuzzeneWaveBeaming,Langley1996,LangleyBeamGrillage}. 
Devices that combine, for example, beams and plates with additional resonating structures have been used to produce novel effects associated with bending and rotation, leading to phononic properties not typically found in optical or plasmonic materials, such as uni-directional wave modes, allowing the transmission and localisation of energy~\cite{ChiralWaves,ParallelGyro,GregEnergyHarvest,DeflectElasticPrism,Ayz-StepResonantPrimitive,NievesNon-ReciprocGyroLat}.
The Green's function is the canonical object of study for many problems associated with wave propagation in structured solids as it contains all the fundamental information (such as the dispersive properties) corresponding to the dynamic response of the system. 
There is a substantial amount of literature on Green's functions, particularly for scalar systems, such as the wide variety of topologies and wave propagation modes (free and forced) studied on mass-spring lattices,~\cite{DynamicAnisotropy,MathModelWaves,MartinGFScattering,VanelAsymDynamicGF}. 
Green's functions afford the means to control wave propagation and analyse defects that occur in lattice systems~\cite{SlepyanCrack,ColquittLineOfDefects}.
A method for the localisation of energy in scalar lattices was provided in~\cite{BandGapGFO}, where the Green's function was used to customise evanescent defect modes that coincide with localised band gap modes. 

Considering lattices of beams, various studies have explored systems of Rayleigh beams, including different topologies, the effects of rotational inertia, and the propagation of both in-plane and out-of-plane vibrations~\cite{PiccMovDisperAndLocalis,DispersionDegeneracies,FreeAndForced,FloqBlochWavesRayleigh}.
Forcing in the form of applied moments is uncommon in flexural systems, but has been shown to produce interesting wave patterns with varying degrees of anisotropy~\cite{FreeAndForced}.

Lattices of Euler--Bernoulli beams have been the subject of studies focusing on many different facets, including their topology and propagation modes, and have been combined with additional resonators to produce interesting phenomena that can be used as wave guides~\cite{DynamicAnisotropy, RuzzeneWaveBeaming,FloqBlochWavesRayleigh,RotationalInertiaInterface,ColquittDispersionMicrostructure}. 
One such example~\cite{DispersionDegeneracies} demonstrated how the interface between a half-plane of Euler--Bernoulli beams, and a half-plane of Rayleigh beams can produce negative refraction for incoming lattice waves.
Another example~\cite{ChiralWaves} combined a periodic array of Euler--Bernoulli beams on a plate with gyroscopic resonators, leading to `chiral flexural waves' and one way edge waves at the chiral interface.  
Unlike lattices of Rayleigh beams, rotational inertia is usually neglected on lattices of Euler--Bernoulli beams as standard.
In this study, we explore the effect of rotational inertia and torsional stiffness in lattices of Euler--Bernoulli beams with remarkable results.
We emphasise that torsional interactions are often neglected when studying flexural systems insofaras the two are treated independently.
However, we show that properly accounting for the interaction between these flexural and torsional effects leads to unprecedented levels of control over the dispersive properties of the lattice.
Alongside these observations, we also compare the different waveforms generated in the lattices from applied forcing, versus applied moments and show that with applied moments, the lattice can act as a wave guide without the need for additional resonators or interfaces. 

The structure of the paper is as follows: firstly, in \S~\ref{sec2DLattice} we construct a class of Green's functions for a 2D square lattice of Euler--Bernoulli beams, where the nodes have mass and rotational inertia and the beams possess torsional stiffness.
In \S~\ref{ComsolMult}, we use applied forces and moments to induce dynamic anisotropy, including unusual asymmetric propagative (non-reciprocal) modes.
In \S~\ref{Sec2DRotAdnTorsion} we show that altering the magnitude of the rotational inertia and torsional stiffness provides a range of possible (and significantly different) dispersion diagrams.
In \S~\ref{BeamChain} we study the related problem of a 1D chain of Euler--Bernoulli beams with nodes that possess rotational inertia, and show that the absence of torsional interactions allows us to evaluate the full class of Green's functions for the chain in closed form.
Lastly, in \S~\ref{setup} we generalise the approach used in the previous sections to provide a method for constructing Green's functions on discrete lattices of any dimension, with lattice connections that take any form, and then go on to derive custom localised defect modes. 

In this study, the control we demonstrate over the direction of wave propagation lays the foundation for designing metamaterials of Euler--Bernoulli beams that have tailor-engineered properties.

\section{A square lattice of Euler--Bernoulli beams}
\label{sec2DLattice}

\begin{figure}[ht]
    \centering
    \begin{subfigure}[t]{0.3\textwidth}
        \centering
        \includegraphics[width=0.9\textwidth]{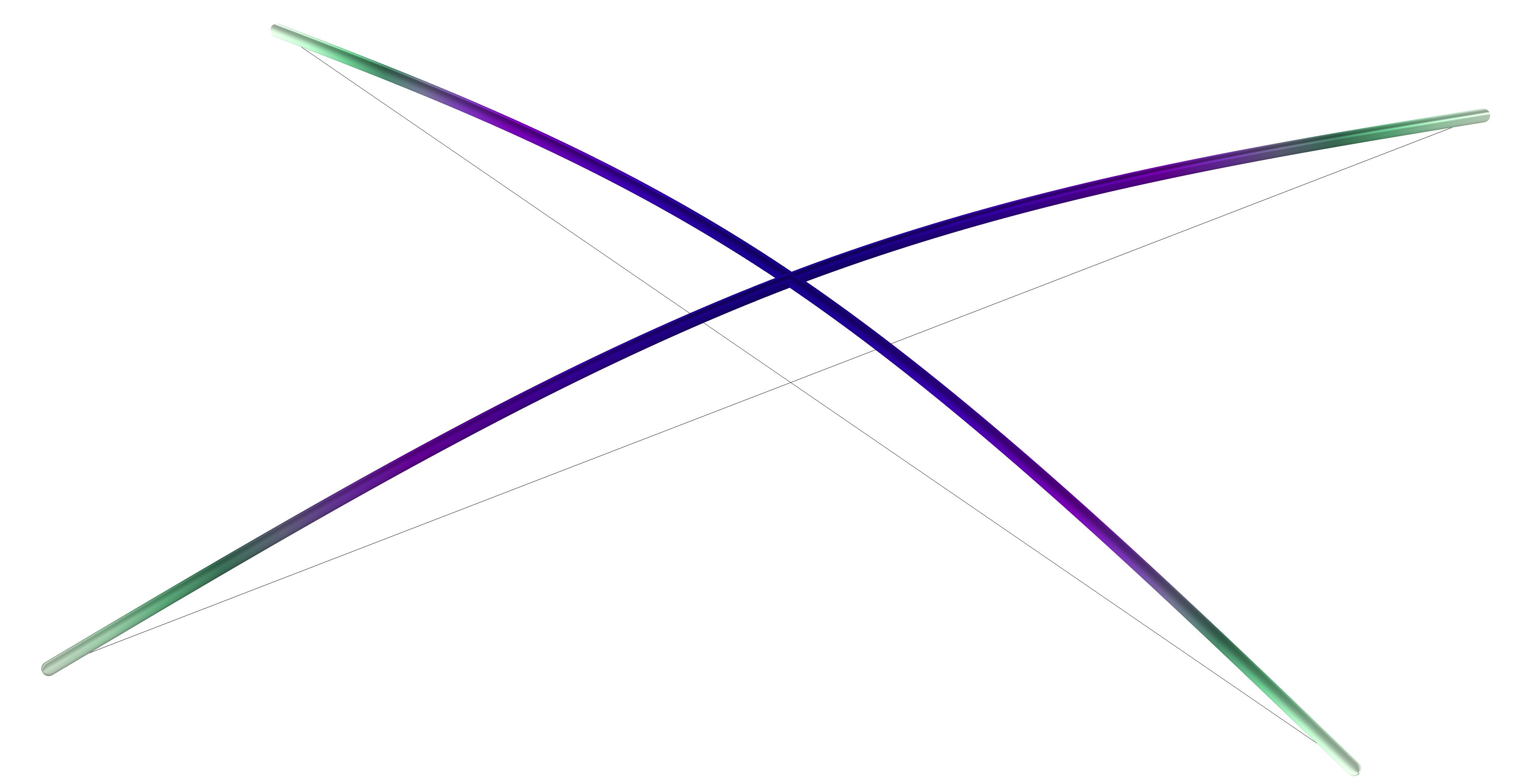}
    \end{subfigure}%
    \hfill
    \begin{subfigure}[t]{0.3\textwidth}
        \centering
        \includegraphics[width=0.9\textwidth]{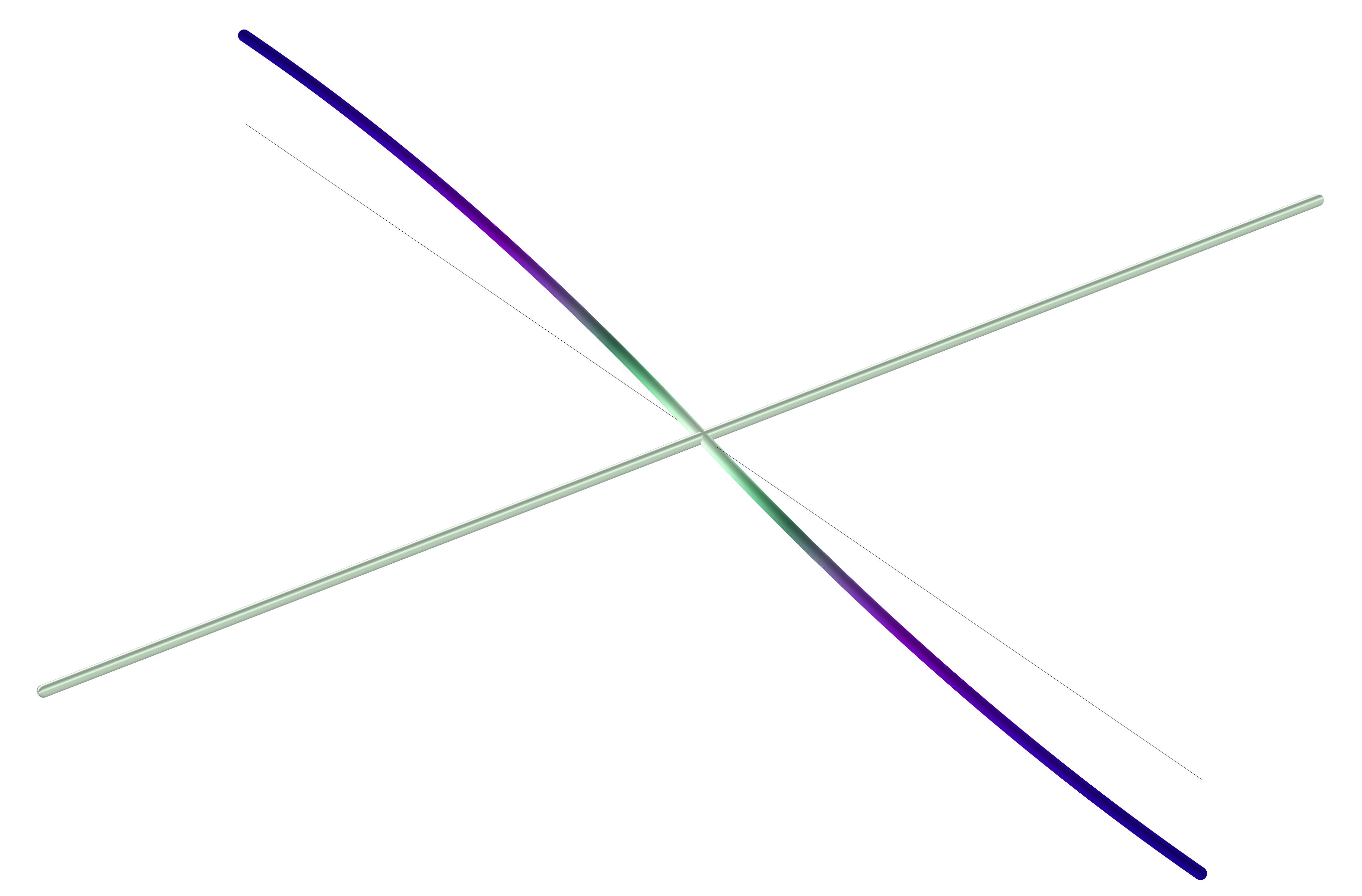}
    \end{subfigure}
    \hfill
    \begin{subfigure}[t]{0.3\textwidth}
        \centering
        \includegraphics[width=0.9\textwidth]{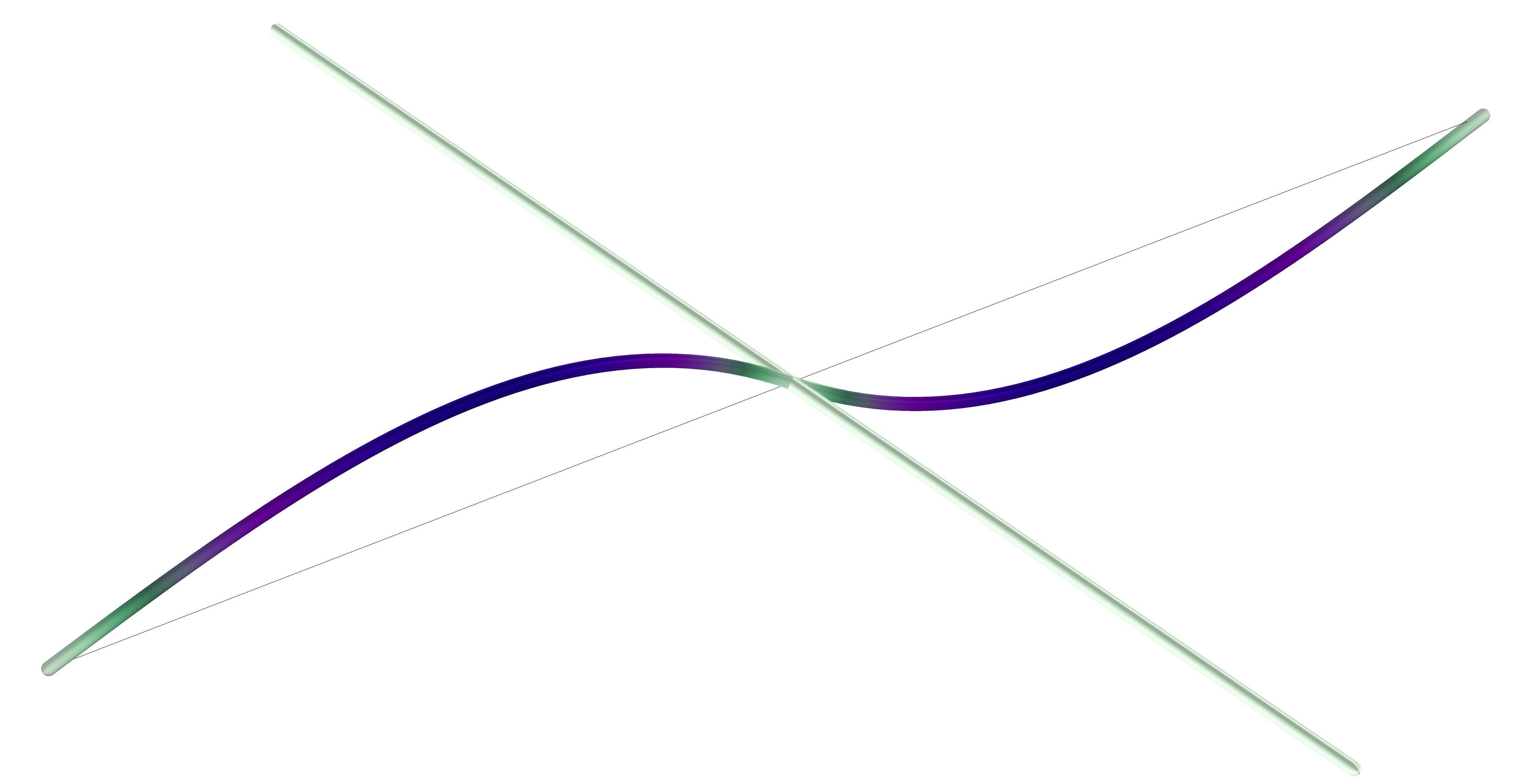}
    \end{subfigure}
    \caption{The eigenmodes of the lattice unit cell, illustrating how out-of-plane flexural deformations couple with torsional displacements at the junctions.}
    \label{COMeigenmodes}
\end{figure}

In this section we consider an infinite square lattice of thin Euler--Bernoulli beams with junctions possessing both mass and rotational inertia. 
The beams have both flexural rigidity and torsional stiffness, meaning that the torsional rotation and flexural displacement at the junctions are coupled.
In particular, whilst the flexural displacement and rotation within a beam are coupled, the torsional rotation is decoupled within the beam itself.
At the junctions, however, the flexural rotation in a beam will induce a torsional rotation in the perpendicular beams, coupling the two classes of waves.
This coupling of flexural and torsional interactions is often overlooked in flexural systems but, as will be shown later (\S~\ref{ComsolMult}), properly accounting for these interactions leads to novel and interesting effects.
In Figure~\ref{COMeigenmodes}, the eigenmodes of the lattice are illustrated using a finite element model of the lattice unit cell.

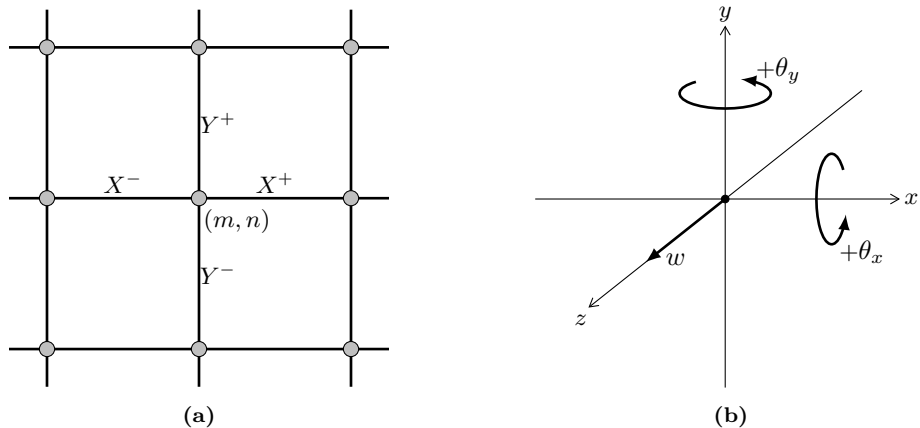
\begin{figure}[ht]
    \centering
    \begin{subfigure}[b]{0.5\textwidth}
        \centering
        \begin{tikzpicture}
\draw[line width=1, color=black] (-0.5,0) -- (4.5,0);
\draw[line width=1, color=black] (-0.5,-2) -- (4.5,-2);
\draw[line width=1, color=black] (-0.5,2) -- (4.5,2);
\node[] at (3,0.2) {\small $X^{+}$};
\node[] at (1,0.2) {\small$X^{-}$};
\node[] at (2.25,1) {\small$Y^{+}$};
\node[] at (2.25,-1) {\small$Y^{-}$};
\node[] at (2.5,-0.3) {\small $(m,n)$};
\draw[line width=1, rotate around={90:(0,0)}, color=black] (-2.5,0) -- (2.5,0);	
\begin{scope}[shift={(2,0)}]
\draw[line width=1, rotate around={90:(0,0)}, color=black] (-2.5,0) -- (2.5,0);
\end{scope}	
\begin{scope}[shift={(4,0)}]
\draw[line width=1, rotate around={90:(0,0)}, color=black] (-2.5,0) -- (2.5,0);
\end{scope}	
\draw[fill=lightgray] (0,0) circle (0.1);
\draw[fill=lightgray] (2,0) circle (0.1);
\draw[fill=lightgray] (4,0) circle (0.1);
\draw[fill=lightgray] (0,2) circle (0.1);
\draw[fill=lightgray] (2,2) circle (0.1);
\draw[fill=lightgray] (4,2) circle (0.1);
\begin{scope}[shift={(0,-5)}]
\draw[fill=lightgray] (0,3) circle (0.1);
\draw[fill=lightgray] (2,3) circle (0.1);
\draw[fill=lightgray] (4,3) circle (0.1);
\end{scope}
\end{tikzpicture}
        \caption{}
        \label{LatticeAndAxesLattice}
    \end{subfigure}%
    \hfill
    \begin{subfigure}[b]{0.5\textwidth}
        \centering
        \begin{tikzpicture}
\draw[->, >=angle 60] (-2.5,0) -- (2.3,0);
\draw[->, >=angle 60] (0,-2.5) -- (0,2.3);
\draw[<-, >=angle 60] (-1.8,-1.44) -- (1.8,1.44);
\draw[<-, >=latex, line width=1] (-1.05,-0.84) -- (0,0);
\node[] at (2.45,0)  {\small $x$};
\node[] at (0,2.45)  {\small $y$};
\node[] at (-1.9,-1.6)  {\small $z$};
\node[] at (-0.65,-0.8)  { $w$};
\node[] at (1.8,-0.7)  {{\small{+}}$\theta_{x}$};
\node[] at (0.7,1.7)  { {\small{+}}$\theta_{y}$};
\draw[fill=black] (0,0) circle (0.05);
\tikzset{
    partial ellipse/.style args={#1:#2:#3}{
        insert path={+ (#1:#3) arc (#1:#2:#3)}    }}
\draw[->, >=latex,line width=1] (1.4,0) [partial ellipse=40:340:0.2 and 0.6];
\draw[->, >=latex,line width=1] (0,1.4) [partial ellipse=-230:70:0.6 and 0.2];
\end{tikzpicture}
       \caption{}
               \label{LatticeAndAxesAxes}
    \end{subfigure}
        \caption{(a) The Euler--Bernoulli beam lattice geometry, with (b) the corresponding coordinate axes and displacements.}
        \label{LatticeAndAxes}
\end{figure}

We construct the lattice in the $xy$-plane and therefore define the out-of-plane translational motion along the $z$-axis.
For convenience, we choose the mass of the nodes, the length of the beams and the flexural rigidity as natural units.
Each lattice node is labelled by the double-index $(m,n)\in\mathbb{Z}^2$, such that $\vec{0}$ denotes the node located at $(x,y) = (0,0)$.
We introduce the generalised displacement vector $\vec{u}_{(m,n)}=[w(m,n),\theta_{x}(m,n),\theta_{y}(m,n)]^{\mathrm{T}}$ to describe the displacement and rotation of each node. 
The first component of $\vec{u}_{(m,n)}$ describes the translation $w(m,n)$ of the $(m,n)^\text{th}$ node, while the second and third components, $\theta_{x}(m,n)$ and $\theta_{y}(m,n)$, characterise the rotational displacement of the nodes around the respective coordinate axes, as indicated in Figure~\ref{LatticeAndAxesAxes}.

The four beams connected at the node $(m,n)$ are denoted by $X^{+}$, $X^{-}$, $Y^{+}$ and $Y^{-}$ corresponding to the axes on which the beams lie.
Hence, the beam spanning $m\leq x \leq m+1$ is denoted $X^{+}$ and so forth, as shown in Figure~\ref{LatticeAndAxesLattice}.
The out-of-plane flexural deformation $w(x,y)$ of the lattice links is governed by the fourth order differential equation for Euler--Bernoulli beams,
 \begin{equation}
    \pdv[4]{w}{x} = 0,\;\text{on}\;X^{\pm} \qquad \text{and} \qquad
    \pdv[4]{w}{y} = 0,\;\text{on}\;Y^{\pm}.
    \label{2Deq4thorderbeam}
\end{equation}
The first-order spatial derivatives $w_{,x}(x,y)$, $w_{,y}(x,y)$, give the angles of flexural rotation, about the $x$- and $y$-axes respectively, associated with the flexural deformation; here subscript letters proceeded by commas indicate partial derivatives with respect to the relevant spatial variables. 
We define $\tau_{x} (x,y)$ as the torsion angle that the $X^{\pm}$ beams experiences about the $x$-axis, and likewise for $\tau_{y} (x,y)$ about the $y$-axis. 
The torsion angles each satisfy
\begin{equation}
    \pdv[2]{\tau_{x}}{x}=0 \qquad \text{and} \qquad \pdv[2]{\tau_{y}}{y}=0.
    \label{2Dtrosionangles}
\end{equation}

To construct the equation of motion, we first consider the forces and bending moments that the $X^{\pm}$ beams apply to the node $(m,n)$. 
From equations~\eqref{2Deq4thorderbeam} and~\eqref{2Dtrosionangles}, we see that the flexural deformation and torsion angles are cubic and linear polynomials respectively. 
The coefficients of these polynomials are found by imposing boundary conditions at the ends of the $X^{\pm}$ beams; we use the following constants,
\begin{multicols}{3}
\noindent
\begin{align*}
   w(m-1,n) &= W_{1}\\ 
   w_{,x}(m-1,n) &= \Theta_{y}^{(1)}\\
   \tau_{x}(m-1,n) &= -\Theta_{x}^{(1)}
   \end{align*}
\begin{align*}
   w(m,n) &= W_{0} \\
   w_{,x}(m,n) &= \Theta_{y}^{(0)}\\
   \tau_{x}(m,n) &= -\Theta_{x}^{(0)}
   \end{align*}
\begin{align*}
   w(m+1,n) &= W_{1}\\
   w_{,x}(m+1,n)  &= \Theta_{y}^{(1)}\\
   \tau_{x}(m+1,n) &= -\Theta_{x}^{(1)}.
   \end{align*}
\end{multicols}
\noindent
Careful attention is required when considering the directions of the flexural moments and torsional moments around the in-plane coordinate axes, shown in Figure~\ref{LatticeAndAxesAxes}.
We stipulate that positive $\Theta_{x}$ refers to an anticlockwise angle around the $x$-axis, and likewise for the $y$-axis.
Using the boundary conditions, we arrive at expressions for the flexural displacement of the $X^{+}$ beam
\begin{equation}
     w (x,n)_{X^{+}} = (\Theta_{y}^{(1)} - 2 W_{1} + \Theta_{y}^{(0)} + 2 W_{0})\,x^{3} + (3 W_{1} - \Theta_{y}^{(1)} - 2 \Theta_{y}^{(0)} - 3 W_{0})\,x^{2} + \Theta_{y}^{(0)}x + W_{0},
     \label{2DX+polynomialW}
\end{equation}
and the torsion angle of the $X^{+}$ beam
\begin{equation}
    {\tau_{x}(x,n)}_{X^{+}} = (\Theta_{x}^{(0)} - \Theta_{x}^{(1)})\,x + \Theta_{x}^{(0)}.
    \label{2DX+polynomialTors}
    \end{equation}
We also find the equivalent expressions for the $X^{-}$ beam, $w (x,n)_{X^{-}}$ and ${\tau_{x}(x,n)}_{X^{-}}$.

The shear force, the flexural bending moment and the torsional moment induced in the $X^{\pm}$ beams are
\begin{equation}
F(x,n) = - \pdv[3]{w}{x} ,
\quad
M_{flex}(x,n) = \pdv[2]{w}{x}
\quad\text{and}\quad  M_{tors}(x,n)= - C \pdv{\tau_{x}}{x} ,
   \label{2DeqForceMoments}
\end{equation}
respectively, where $C$ in $M_{tors}$ is the non-dimensionalised torsional stiffness coefficient and the negative sign corresponds to the direction of the moment about the $x$-axis~\cite{Graff}.
The forces and moments that a beam applies to the node $(m,n)$ can be expressed in terms of the generalised forcing vector $\mathbb{F}_{(m,n)} = [ F(m,n) ,  \Phi_{x}(m,n), \Phi_{y}(m,n)]^\mathrm{T}$; where the moments $\Phi_{x}(m,n)$ and $\Phi_{y}(m,n)$ describe the total bending moment about their respective axes at node $(m,n)$, and so each contains both flexural moments $M_{flex}$ and torsion $M_{tors}$.
As such, the forces and moments that the $X^{\pm}$ beams apply to the node $(m,n)$ are
\begin{equation}
\begin{aligned}
\begin{bmatrix} 
F(m,n)\\
\Phi_{x}(m,n) \\
\Phi_{y}(m,n)
\end{bmatrix}_{(X^{+})}
&=
\begin{bmatrix} 
-12 & 0 & -6\\
0 & -C & 0\\
-6 & 0 & -4
\end{bmatrix}_{(A)}
\begin{bmatrix} 
W_{0} \\
\Theta_{x}^{(0)}\\
\Theta_{y}^{(0)}
\end{bmatrix}
+
\begin{bmatrix} 
12 & 0 & -6\\
0 & C & 0\\
6 & 0 & -2
\end{bmatrix}_{(B)}
\begin{bmatrix} 
W_{1} \\
\Theta_{x}^{(1)}\\
\Theta_{y}^{(1)}
\end{bmatrix}; \\
\begin{bmatrix} 
F(m,n) \\
\Phi_{x}(m,n)\\
\Phi_{y}(m,n)
\end{bmatrix}_{(X^{-})}
&=
\begin{bmatrix} 
-12 & 0 & 6\\
0 & -C & 0\\
6 & 0 & -4
\end{bmatrix}_{(A)}
\begin{bmatrix} 
W_{0} \\
\Theta_{x}^{(0)}\\
\Theta_{y}^{(0)}
\end{bmatrix}
+
\begin{bmatrix} 
12 & 0 & 6\\
0 & C & 0\\
-6 & 0 & -2
\end{bmatrix}_{(B)}
\begin{bmatrix} 
W_{1} \\
\Theta_{x}^{(1)}\\
\Theta_{y}^{(1)}
\end{bmatrix};
\end{aligned}
\label{2DeqX+-forces}
\end{equation}
and we label the matrices for each beam $X^{+}_A$ and $X^{+}_{B}$, $X^{-}_{A}$ and $X^{-}_{B}$, as indicated in equation~\eqref{2DeqX+-forces}.

Following the same method, we apply boundary conditions to the ends of the $Y^{\pm}$ beams;
since flexion in the $X^{\pm}$ beams induces torsion in the $Y^{\pm}$ beams, this is reflected in the continuity condition at $(m,n)$ such that $w_{,x}(m,n)=\Theta_{y}^{(0)}=\tau_{y}(m,n)$ and $w_{,y}(m,n)=-\Theta_{x}^{(0)}=\tau_{x}(m,n)$.
Using the boundary conditions, the polynomials for the flexural displacement $w (m,y)_{Y^{\pm}}$ and torsion angles ${\tau_{y}(m,y)}_{Y^{\pm}}$ associated with the $Y^{\pm}$ beams are found. 
The forces and bending moments that the $Y^{\pm}$ beams apply to the node $(m,n)$ are then found using the $y$-axis equivalent expressions to equation~\eqref{2DeqForceMoments}. 
In parallel with equation~\eqref{2DeqX+-forces}, we express the forces and moments from the $Y^{\pm}$ beams as vectors, and we label the corresponding matrices $Y^{+}_{A}$ and $Y^{+}_{B}$,  $Y^{-}_{A}$ and $Y^{-}_{B}$.

Combining the forces and moments in both the $x$- and $y$-directions with Newton's second law for a time-harmonic system, the equation of motion of the $(m,n)^\mathrm{th}$ node is then
\begin{multline}
    -\omega^{2} \mathsf{M} \vec{u}_{(m,n)} =\big[ X^{+}_{A} +Y^{+}_{A}+X^{-}_{A}+Y^{-}_{A}\big] \vec{u}_{(m,n)}\\
    + X^{+}_{B} \vec{u}_{{(m+1,n)}} + Y^{+}_{B} \vec{u}_{{(m,n+1)}} + X^{-}_{B} \vec{u}_{{(m-1,n)}} + Y^{-}_{B} \vec{u}_{{(m,n-1)}},   
    \label{eqnthunitcelllattice}
\end{multline}
where $\vec{u}_{{(m,n)}}$ is again the generalised displacement vector of the node $(m,n)$ and the $3 \times 3$ matrix $\mathsf{M} = \diag[1,\mu,\mu]$ describes the inertial properties of the lattice. 
The first component of $\mathsf{M}$ corresponds to the (unit) mass of the junction associated with the translational inertia, whilst the second and third components correspond to the rotational inertia associated with the flexural and torsional deformations, which we denote $\mu$.
We apply the discrete Fourier transformation 
\begin{equation}
\vec{u}^{F}(k_{1},k_{2}) = \sum_{(m,n)\in\mathbb{Z}^2} \mathrm{e}^{-\I k_{1}m} \mathrm{e}^{-\I k_{2}n}\vec{u}_{(m,n)} \, ,
\label{eq2dFT}
\end{equation}
to equation~\eqref{eqnthunitcelllattice}, where $\vec{u}^{F}(k_{1},k_{2}) = [{W^{F}}(k_{1},k_{2}), {\Theta_{x}}^{F}(k_{1},k_{2}), {\Theta_{y}}^{F}(k_{1},k_{2})]^{\mathrm{T}}$ is the generalised displacement in reciprocal space in terms of the spectral parameters $k_{1}$ and $k_{2}$.
Following the Fourier transformation, we arrive at the lattice's equation of motion in reciprocal space,
\begin{multline}
   0 =  \bigg[\omega^{2} \mathsf{M} +X^{+}_{A} + Y^{+}_{A}+ X^{-}_{A} + Y^{-}_{A}
   + \mathrm{e}^{\I k_{1}}X^{+}_{B} +\mathrm{e}^{\I k_{2}}Y^{+}_{B} \\
   + \mathrm{e}^{-\I k_{1}}X^{-}_{B} + \mathrm{e}^{-\I k_{2}}Y^{-}_{B}\bigg] \vec{u}^{F}(k_{1},k_{2}).
    \label{eqlatticeunitcellFD}
\end{multline}
As such, the three components of $\vec{u}^{F}(k_{1},k_{2})$ are defined as the Fourier transformed real space displacements $w(m,n)$, $\theta_{x}(m,n)$ and $\theta_{y}(m,n)$, respectively. 

\subsection{The dispersion equation} \label{Sec2DdispersionSurfaces}

The dispersion equation $\sigma(\omega, k_{1}, k_{2}) = 0$ arises from the solvability condition of equation~\eqref{eqlatticeunitcellFD}, where
\begin{multline}
\sigma(\omega, k_{1}, k_{2}) =  144\sin ^2(k_{1})\zeta(k_{1}, k_{2}) + 144 \sin ^2(k_{2} )\zeta(k_{2}, k_{1}) \\
+ \left(24 \cos(k_{2} )+24 \cos (k_{1})+\omega ^2-48\right)\zeta(k_{1}, k_{2}) \zeta(k_{2}, k_{1}),
\label{eq2Ddispersioneq}
\end{multline}
and we define the repeated function
\begin{equation}
    \zeta (P, Q) = \left(-2 C \cos (P)+ 2 C +4 \cos (Q) -\mu  \omega ^2+8\right).
    \label{2Dzeta}
\end{equation}
As one would expect for a symmetric square lattice, the dispersion equation is symmetric in the spectral parameters $k_{1}$ and $k_{2}$.
We also see that the  dispersion equation is cubic in $\omega^{2}$ and therefore, closed form solutions can be found but are omitted here for brevity.
In addition to the frequency and spectral parameters, $\sigma(\omega, k_{1}, k_{2})$ is dependent on the material constants $\mu$ and $C$, which describe the non-dimensionalised rotational inertia and torsional stiffness, respectively.
In \S~\ref{Sec2DRotAdnTorsion} we investigate the effects of changing $\mu$ and $C$
on the dispersion equation and show that it is possible to produce completely different dispersion diagrams, with and without the existence of a finite band gap. 
This control over the lattice's propagating frequencies (with their preferential directions and subsequent anisotropy) is a particularly interesting feature that provides applications for metamaterials designed to control the propagation of waves in structures. 
\begin{figure}[ht]
\centering
\begin{minipage}{\textwidth}
\centering
\includegraphics[width=0.7\textwidth]{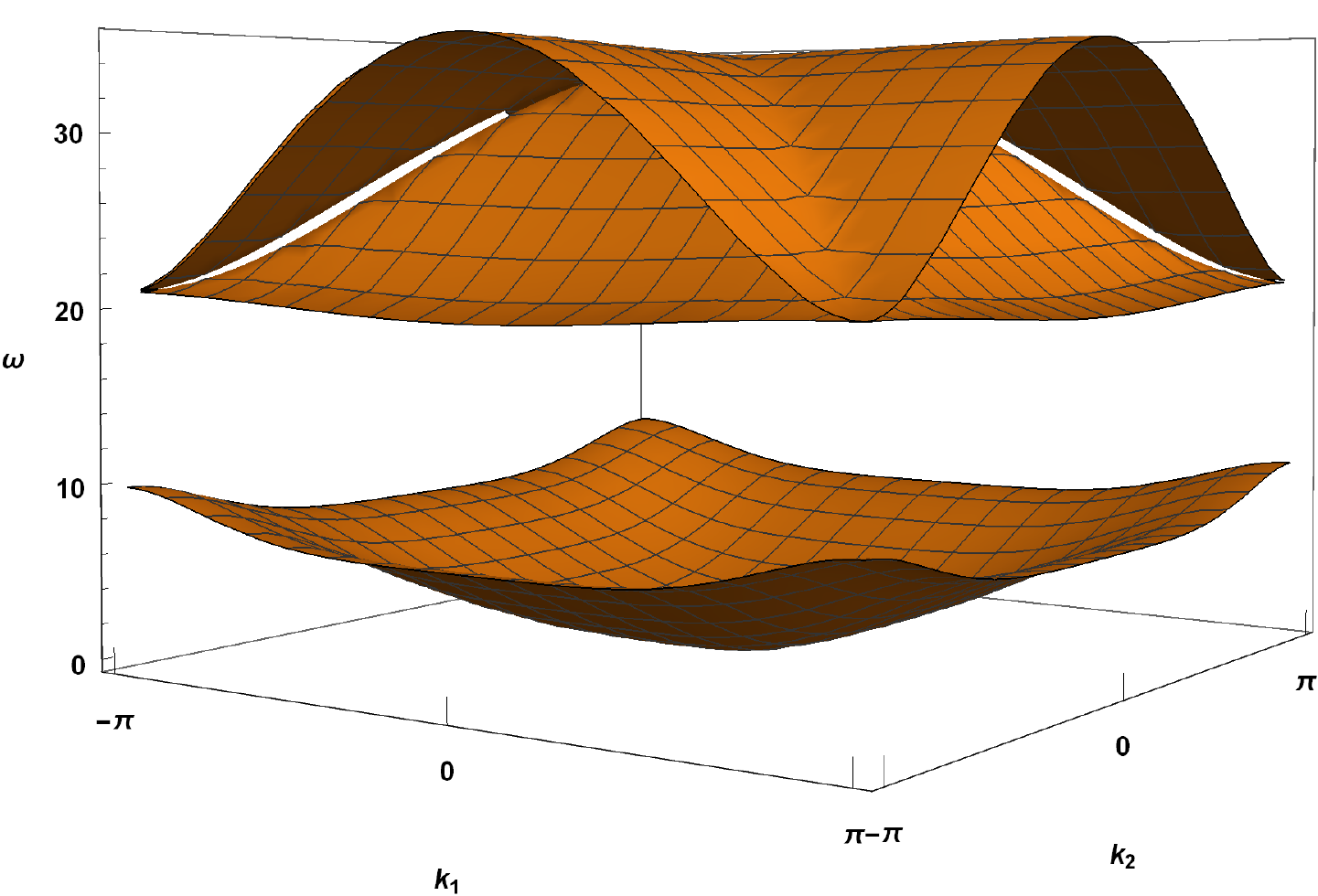}
    \caption{The dispersion diagram for the Euler--Bernoulli beam lattice across the Brillouin zone $( k_{1}, k_{2}) \in [-\pi,\pi]^2$, for the values of $\mu = 0.01$, $C = 0.1$.}
    \label{figlatticedispwithbandgap}
    \end{minipage} 
\end{figure}

In Figure~\ref{figlatticedispwithbandgap}, we plot the solutions to $\sigma(\omega, k_{1}, k_{2}) = 0$ for the illustrative values of $\mu = 0.01$ and $C = 0.1$. 
It is shown that these values of rotational inertia and torsional stiffness produce a dispersion diagram with two distinct pass bands and two band gaps, one finite and also the semi-infinite band gap associated with discrete systems. 
Given that the closed form solutions to the dispersion equation are in-hand, we can determine the exact limits of the band gaps for any desired $\mu$ and $C$.

\subsection{Constructing the Green's function}\label{sec2dvertdisp}

A notable feature of the flexural lattice is the ability to induce translational displacements, rotational moments and combinations thereof.
To construct the Green's function, we consider the application of time-harmonic unit forces and unit moments to the central node of the lattice through the forcing vector $\vec{f}\in \mathbb{C}^{3}$.
In parallel with the components of $\mathbb{F}_{(m,n)}$, the forcing vector $\vec{f}=[f_{w}, f_{\theta_{x}}, f_{\theta_{y}}]^{\mathrm{T}}$ has components corresponding to the application of translational force $f_{w}$ out-of-plane (along the $z$-axis) and two moments $f_{\theta_{x}}$ and $f_{\theta_{y}}$ about the $x$- and $y$-axes respectively. 
For any chosen frequency, there are seven different forcing configurations.
These modes are generated by the permutations of $f_{w}, f_{\theta_{x}}, f_{\theta_{y}} \in \{0, -1\}$ with the exclusion of $f_{w}=f_{\theta_{x}}=f_{\theta_{y}}=0$ corresponding to the absence of any applied forces or moments.
We express the Green's function in reciprocal space in the form
\begin{multline}
   \vec{u}^{F}(k_{1},k_{2}) =  \bigg[\omega^{2} \mathsf{M} +X^{+}_{A} +Y^{+}_{A}+X^{-}_{A}+ Y^{-}_{A}\\
   + \mathrm{e}^{\I k_{1}}X^{+}_{B} +\mathrm{e}^{\I k_{2}}Y^{+}_{B}  + \mathrm{e}^{-\I k_{1}}X^{-}_{B} + \mathrm{e}^{-\I k_{2}}Y^{-}_{B}\bigg]^{-1} \vec{f} . 
   \label{eqGeneralGreenFTLattice}
\end{multline}
The inverse of the discrete Fourier transform is then applied to equation~\eqref{eqGeneralGreenFTLattice} to find the Green's function in real space, $\vec{u}_{{(m,n)}}$.
The resulting components of $\vec{u}_{{(m,n)}}$ are written in terms of Fourier integrals and, whilst useful, are cumbersome.
Therefore, in the interests of brevity and clarity of presentation, we will restrict ourselves to consideration of the flexural displacement component $W^{F}(k_{1},k_{2})$ of the Green's function $\vec{u}^{F}(k_{1},k_{2})$ for an arbitrary applied forcing.
In this case, the spectral form of the flexural displacement is,
\begin{multline}
    W^{F}(k_{1},k_{2}) =
    \frac{\zeta(k_{1}, k_{2}) \zeta(k_{2}, k_{1})f_{w}}{\sigma(\omega,k_{1},k_{2})}
    + \frac{12 \I \sin (k_{2}) \zeta(k_{2}, k_{1}) f_{\theta_{x}}}{\sigma(\omega,k_{1},k_{2})}\\
    -\frac{12 \I \sin (k_{1})\zeta(k_{1}, k_{2}) f_{\theta_{y}} }{\sigma(\omega,k_{1},k_{2})},
    \label{eqWFkkappa}
\end{multline}
where the functions $\sigma$ and $\zeta$ are given in equations~\eqref{eq2Ddispersioneq} and~\eqref{2Dzeta} respectively. 
The difference of sign for the coefficients of $f_{\theta_{x}}$ and $f_{\theta_{y}}$ in equation~\eqref{eqWFkkappa} is a consequence of the sign convention adopted for shear forces and moments in equation~\eqref{2DeqForceMoments}. 
As expected for a symmetric lattice, the coefficient of the linear force $f_{w}$ is symmetric in the two spectral parameters.
We also note that the denominator of equation~\eqref{eqWFkkappa} coincides with the dispersion equation, which arises from the inverted operator matrix in equation~\eqref{eqGeneralGreenFTLattice}.

The inverse Fourier transformation is applied to find the displacement in real space as follows, 
\begin{equation}
    w(m,n) = \frac{1}{4\pi^{2}}\int\limits_{-\pi}^\pi \int\limits_{-\pi}^{\pi}W^{F}(k_{1}, k_{2})\, \mathrm{e}^{\I k_{1} m} \, \mathrm{e}^{\I k_{2} n}\, \dd k_{1} \,\dd k_{2} \, .
    \label{eq2DinverseFT}
\end{equation}
Although explicit closed form representations do not exist in general, the integral form of the Green's function is amenable to numerical evaluation.
In the following sections, we use numerical evaluation and finite element models to explore the effects of changing the forcing vector, and also the frequency with which the force is applied.

\subsection{Anisotropic behaviour of localised modes}\label{2DNumerical}

For a lattice with $\mu=0.01$ and $C=0.1$, consistent with the dispersion diagram in Figure~\ref{figlatticedispwithbandgap}, the lower boundary of the finite band gap is $\omega = 4\sqrt{6}$. 
Using equation~\eqref{eq2DinverseFT}, the flexural displacement band gap Green's functions of the lattice nodes, $w(m,n)$ is evaluated numerically for three examples of the seven forcing options, with each localised mode demonstrating a unique shape. 
Across each example in Figure~\ref{F2Dw9,8vert}, we chose the frequency of excitation to be $\omega = 9.8$, such that it lies in the band gap, close to the band-edge; this demonstrates the rapid rate of decay and high degree of localisation associated with band gap Green's function in this regime.

Firstly, we consider out-of-plane forcing $\vec{f} = [-1,0,0]^{\mathrm{T}}$ and plot the flexural displacement in Figures~\ref{F2Dw9,8vertSIDE} and~\ref{F2Dw9,8vertABOVE}. In Figure~\ref{F2Dw9,8vertSIDE}, the decay envelope of the displacement is clear, while Figure~\ref{F2Dw9,8vertABOVE} displays the periodic oscillation of the nodes and the symmetry between the $x$- and $y$-directions. 
In mechanical lattices the preferential direction of wave propagation is most commonly seen along the principle axes of the lattice. 
Here, the localised mode is radially symmetric leading to isotropic behaviour which can be understood from the complex solutions $(k_1,k_2)\in\mathbb{C}^2$ to the dispersion equation, leading to a uniform decay rate in all directions.

When $\vec{f}= \left[0,-1,0\right]^{\mathrm{T}}$, a clockwise moment is induced about the $x$-axis at the $\vec{0}$ node; for this applied moment, the flexural displacement of the lattice nodes has been plotted in Figures~\ref{F2Dw9,8SRside} and~\ref{F2Dw9,8SRAbove}. 
While the displacement is localised radially, the displacement is not localised to the $y$-axis as one might expect for this type of forcing, this is again due to the complex solutions of the dispersion equation in this regime, meaning the mode has no preferential direction of travel. 
This is in particular comparison to Figure~\ref{FigCOMcombinedforceSR} where, for the same applied moment at a pass band frequency, the displacement field of the propagating mode is highly localised along the $y$-axis.
In Figure~\ref{F2Dw9,8SRAbove}, we have indicated the dotted line $(m,0)$ (the $x$-axis) where the lattice nodes do not have out-of-plane displacement but do experience rotational displacement, as expected with this type of forcing.
We note that a moment applied about the $y$-axis $\vec{f}= \left[0,0,-1\right]^{\mathrm{T}}$ produces the same out-of-plane displacement in the lattice under a $\pi/2$ rotation about the $z$-axis.

\begin{figure}[H]
    \centering
    \begin{subfigure}[t]{0.5\textwidth}
        \centering
        \includegraphics[height=4.7cm]{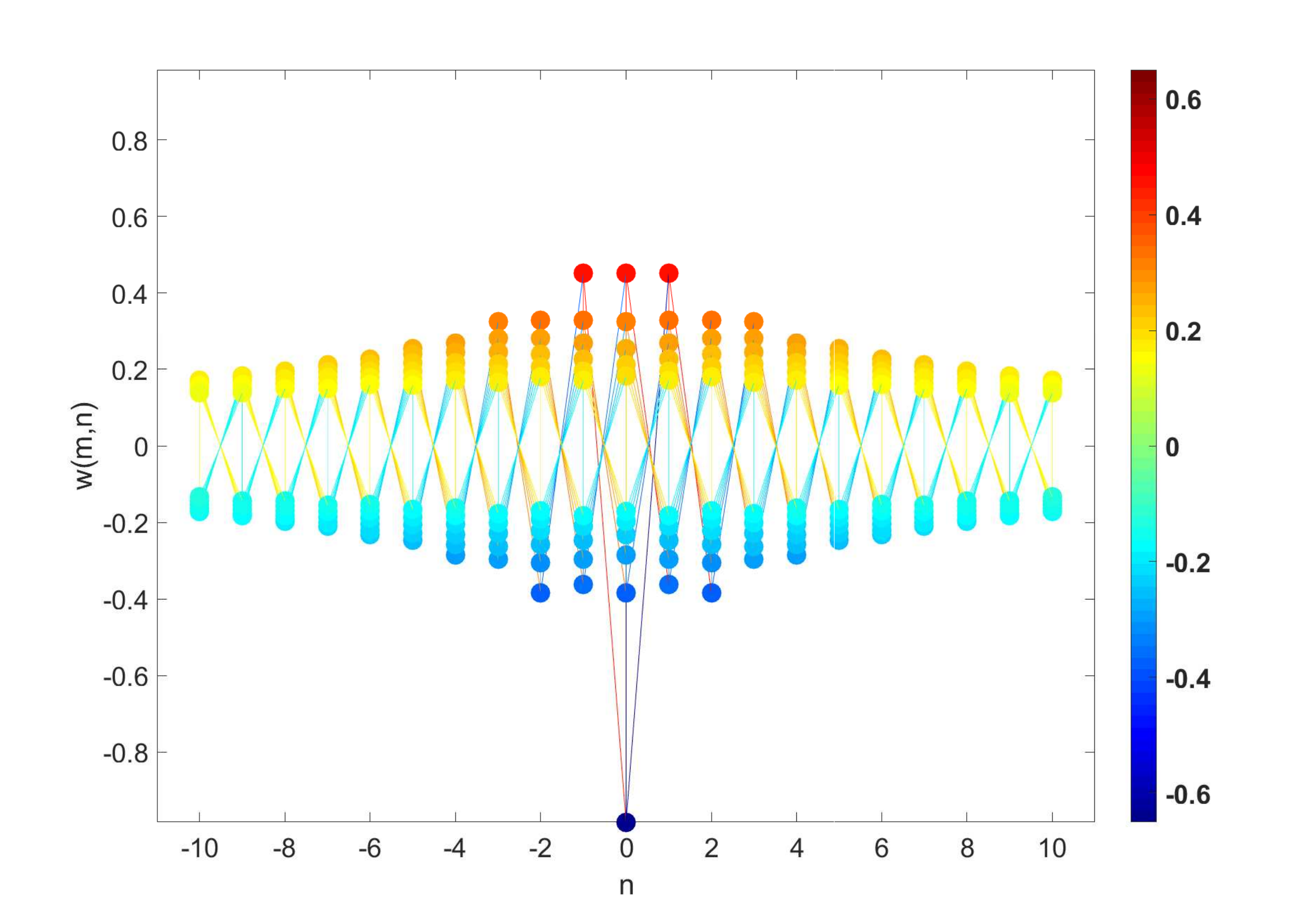}
        \caption{}
        \label{F2Dw9,8vertSIDE}
    \end{subfigure}%
    \hfill
    \begin{subfigure}[t]{0.5\textwidth}
        \centering
        \includegraphics[height=4.7cm]{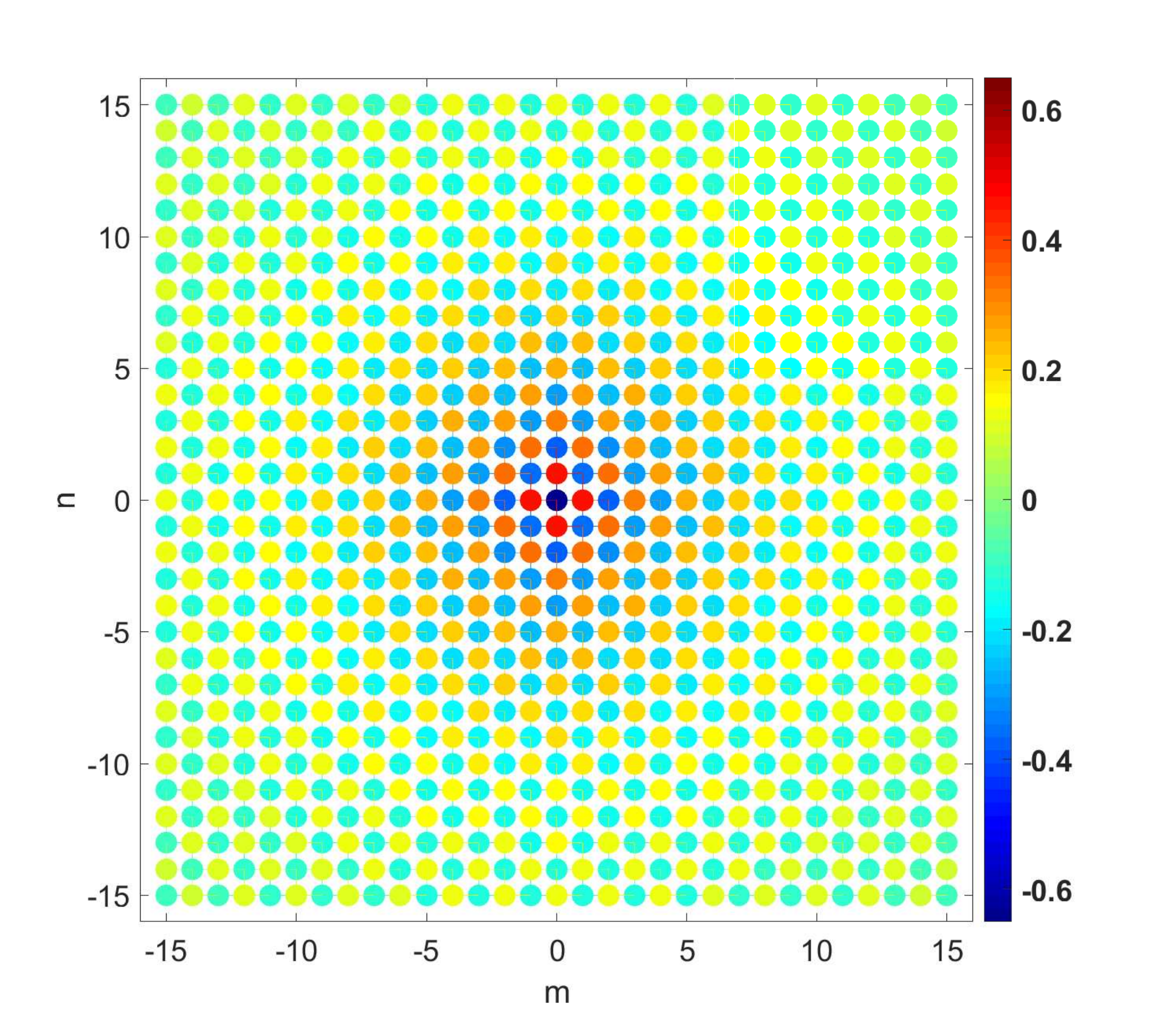}
       \caption{}
       \label{F2Dw9,8vertABOVE}
    \end{subfigure}
    \hfill
    \begin{subfigure}[t]{0.5\textwidth}
        \centering
        \includegraphics[height=4.7cm]{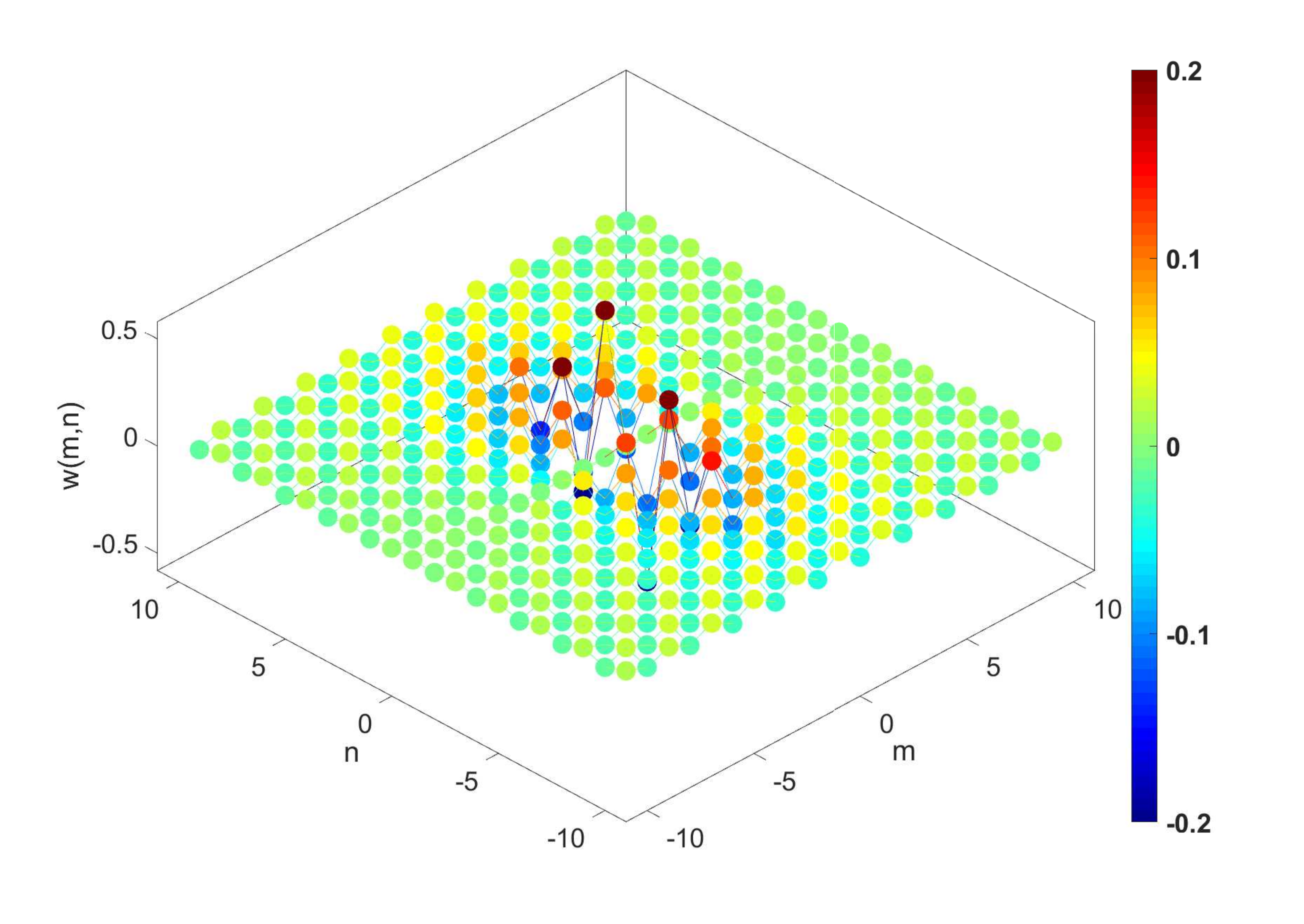}
        \caption{}\label{F2Dw9,8SRside}
    \end{subfigure}%
    \hfill
    \begin{subfigure}[t]{0.5\textwidth}
        \centering
        \begin{tikzpicture}
\node[] at (0,0){\includegraphics[height=4.7cm]{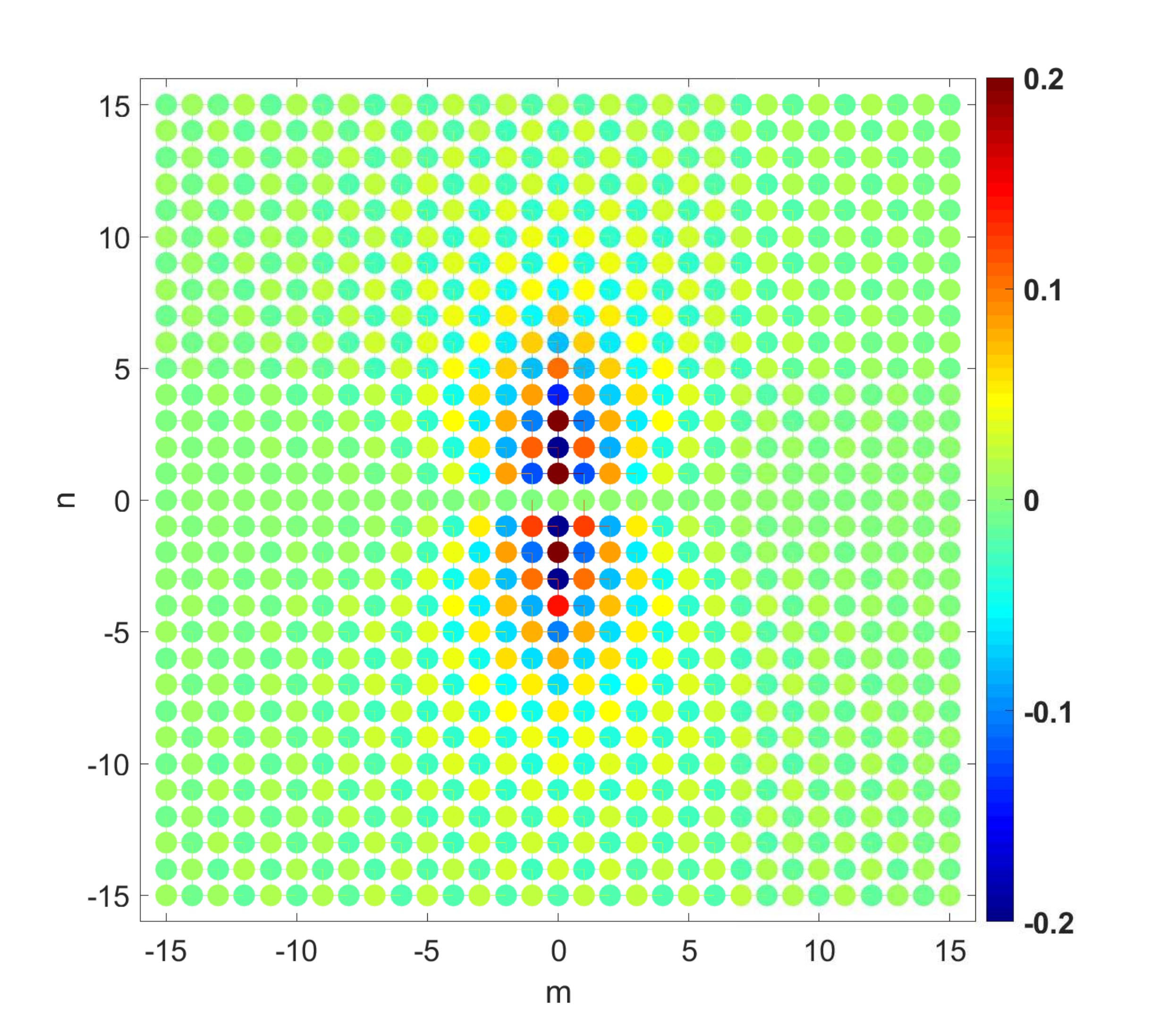}};
\draw[dotted, thick] (-1.9,0.08) -- (1.8,0.08);
        \end{tikzpicture} 
       \caption{}\label{F2Dw9,8SRAbove}
       \end{subfigure}
       \hfill
    \begin{subfigure}[t]{0.5\textwidth}
        \centering
        \includegraphics[height=4.7cm]{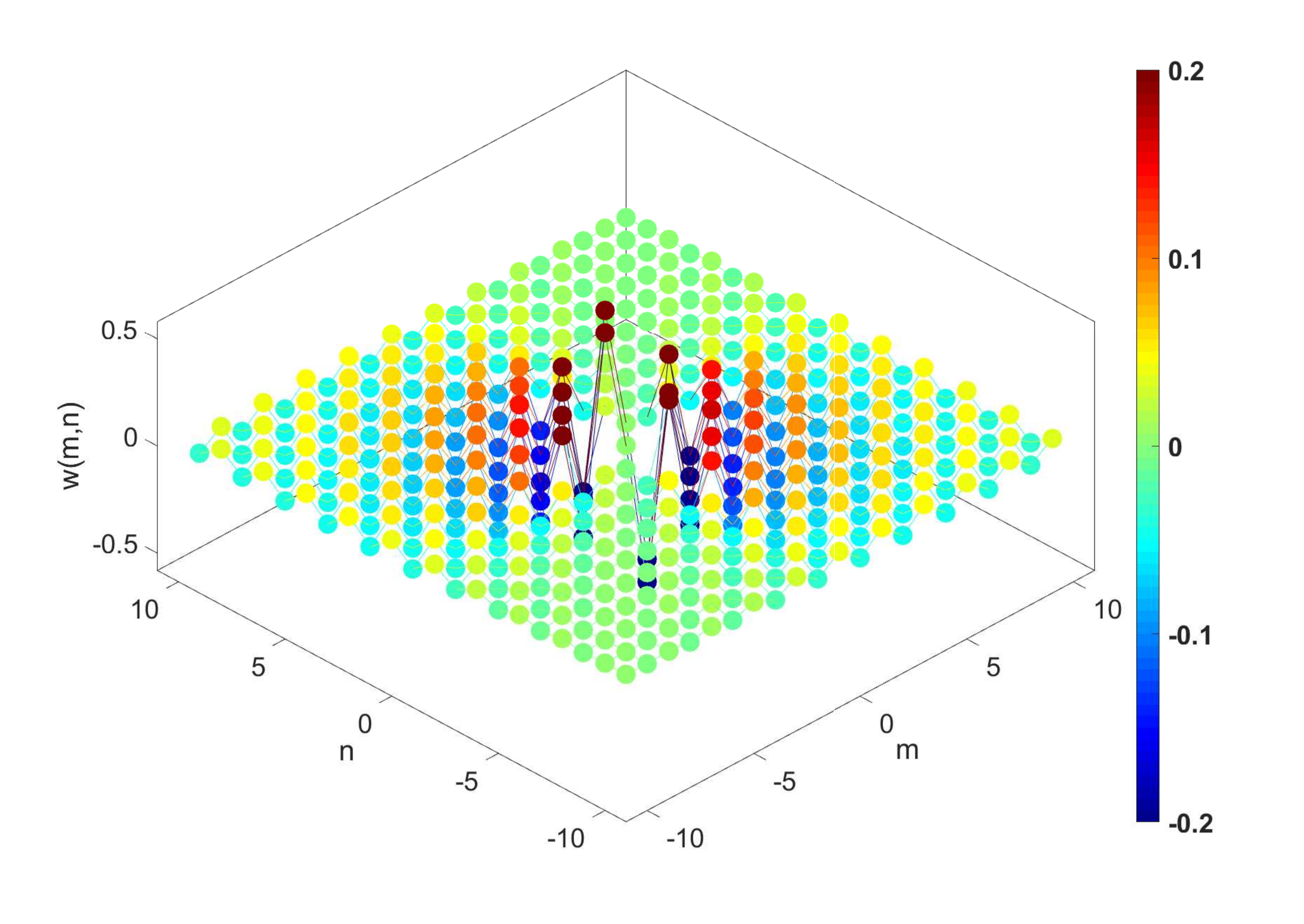}
        \caption{}\label{F2Dw9,8DoubleRside}
    \end{subfigure}%
    \hfill
    \begin{subfigure}[t]{0.5\textwidth}
        \centering
        \begin{tikzpicture}
\node[] at (0,0){\includegraphics[height=4.7cm]{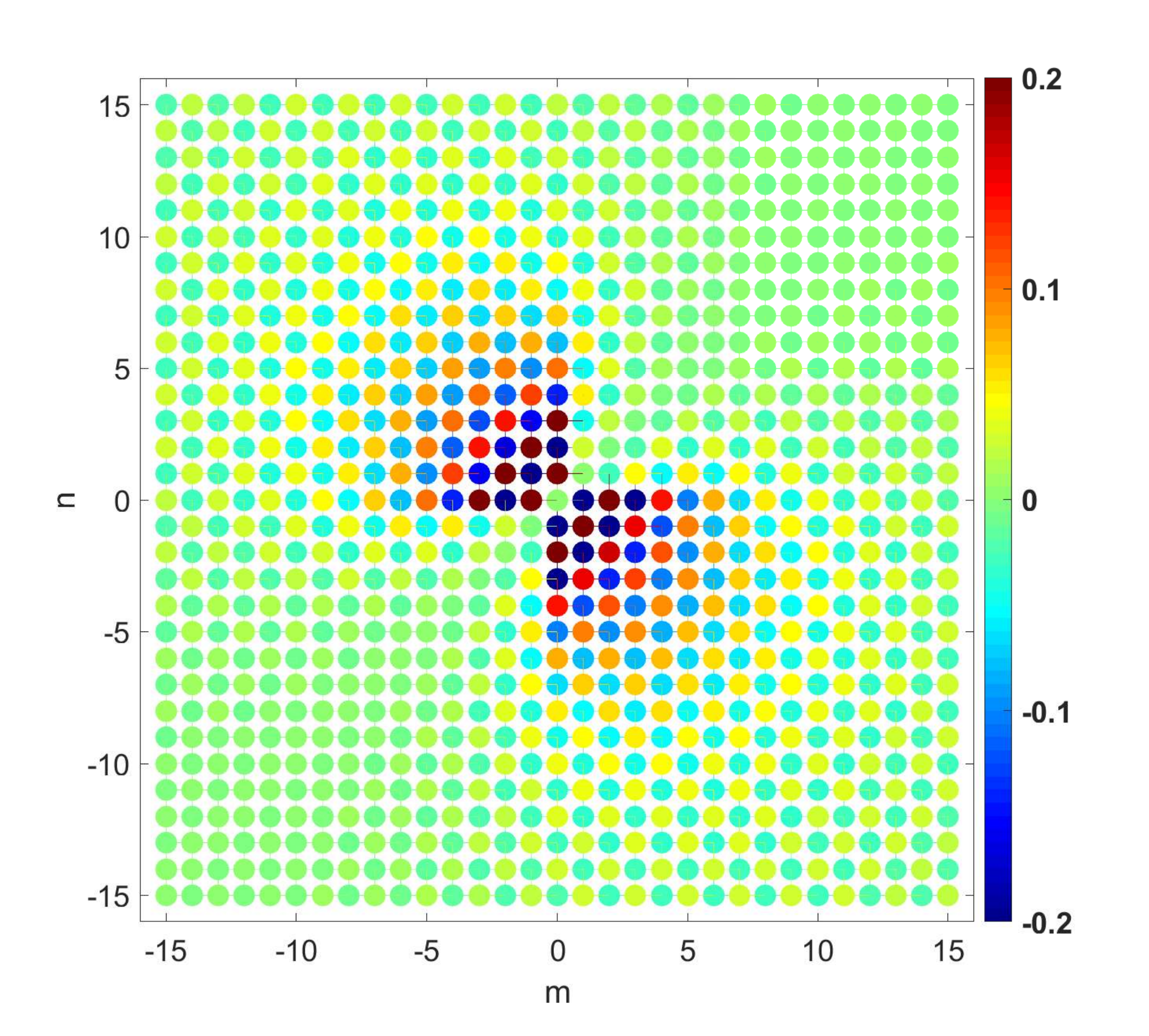}};
\draw[dotted, thick] (-1.88,-1.75) -- (1.75,1.92);
        \end{tikzpicture} 
       \caption{}\label{F2Dw9,8DoubleRAbove}
    \end{subfigure}
    \caption{For a square lattice of Euler--Bernoulli beams, where $\mu = 0.01$, $C = 0.1$, the flexural displacement $w(m,n)$ of the nodes is plotted due to forcing $\vec{f}$ at $\vec{0}$ for $\omega = 9.8$. For $\vec{f} = [-1,0,0]^{\mathrm{T}}$, the lattice is viewed down the $x$-axis (a), and down the positive $z$-axis (b). The forcing $\vec{f} = [0,-1,0]^{\mathrm{T}}$ is demonstrated in (c) and (d); while the forcing $\vec{f} = [0,-1,-1]^{\mathrm{T}}$ is demonstrated in (e) and (f).}
    \label{F2Dw9,8vert}
\end{figure}

With $\vec{f}= \left[0,-1,-1\right]^{\mathrm{T}}$, two unit moments are applied clockwise around the $x$- and $y$-axes at $\vec{0}$ simultaneously and the resulting flexural displacement is plotted in Figures~\ref{F2Dw9,8DoubleRside} and~\ref{F2Dw9,8DoubleRAbove}. 
As with the single moment forcing above, the displacement field generated by the double moments is radially localised but not confined to a single direction; in this case, the line of zero flexural displacement lies on $\pi/4$, as has been indicated by the dotted line.

In addition to evaluating the flexural displacement $w(m,n)$, we can use the inverse Fourier transformation to evaluate the $\theta_{x}(m,n)$ and $\theta_{y}(m,n)$ components of the displacement vector. 
These components give a quantitative measure for the magnitude of the rotation each node experiences around the relevant axis.
In the interests of clarity, the plots of the Green's functions for rotational displacements are omitted here because they do not lend themselves to convenient graphical representation. 
However, it is important to note that these components contribute to the overall lattice response, can be evaluated as described above, and play an important role in the construction of defect modes (cf. \S~\ref{Sec2DmassDefect}).



\subsection{Extreme dynamic anisotropy and non-reciprocity in the pass band}
\label{ComsolMult}

The dispersive behaviour of lattice systems and the associated extreme anisotropy has been well documented and leads to interesting effects such as highly localised waveforms, uni-directional waves, one-way edge modes, and DASERs~\cite{ParabolicMetamaterials,ChiralWaves,DynamicAnisotropy,DeflectElasticPrism,FreeAndForced}.
However these effects are often highly narrowband and usually require careful tuning of the material and geometric parameters through, for instance, the inclusion of resonant elements.
In contrast, here we show how localised waveforms and uni-directional waves are easily achieved in broad frequency regimes as a result of incorporating the additional degrees of freedom associated with the interaction between flexural and torsional motion at the nodes. 
Furthermore, the lattice also supports the highly unusual phenomena of non-reciprocity, despite the lattice being symmetric.

In parallel to the numerical results derived from the analytical Green's function, we develop a novel finite element model to illustrate a range of interesting phenomena in the infinite lattice. 
Using COMSOL Multiphysics\textsuperscript{\textregistered}, we build a lattice model of $201 \times 201$ Euler--Bernoulli beams and impose frequency-independent damping, in the form of complex stiffness values, on the beams in the vicinity of the boundary of the computational window in order to simulate an infinite lattice and minimise any artificial reflections. 
In this section, we set the values of $\mu=0.01$ and $C=0.1$ for the rotational inertia and torsional stiffness respectively across all figures. 


\newlength{\rwa}
\setlength{\rwa}{0.25\textwidth}
\pgfplotsset{scaled ticks = false}
    
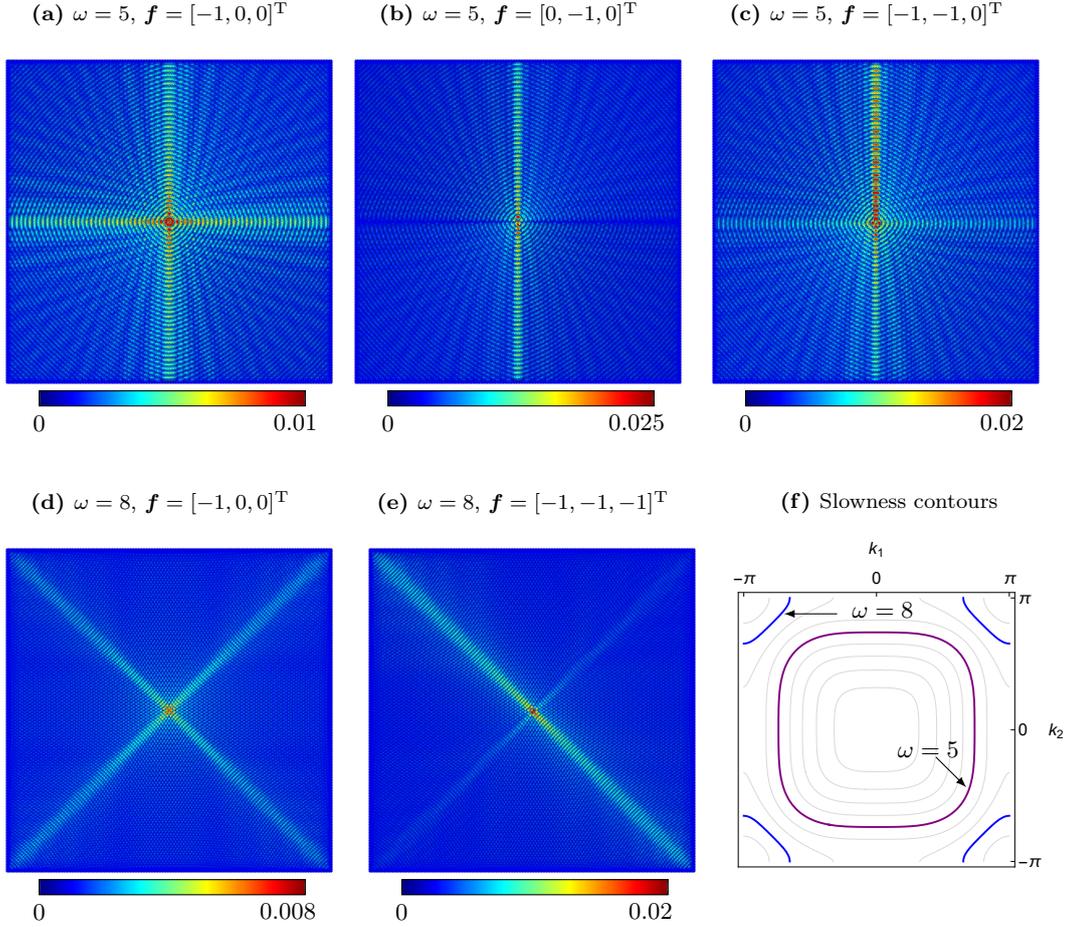
\begin{figure}[ht]
    \centering
     \begin{subfigure}[t]{0.32\textwidth}
        \centering
        \caption{$\omega=5$, $\vec{f} = [-1,0,0]^{\mathrm{T}}$ }
       \label{COMandSLOWw5Pic}
        \begin{tikzpicture}
\node(a){\includegraphics[width=\textwidth]{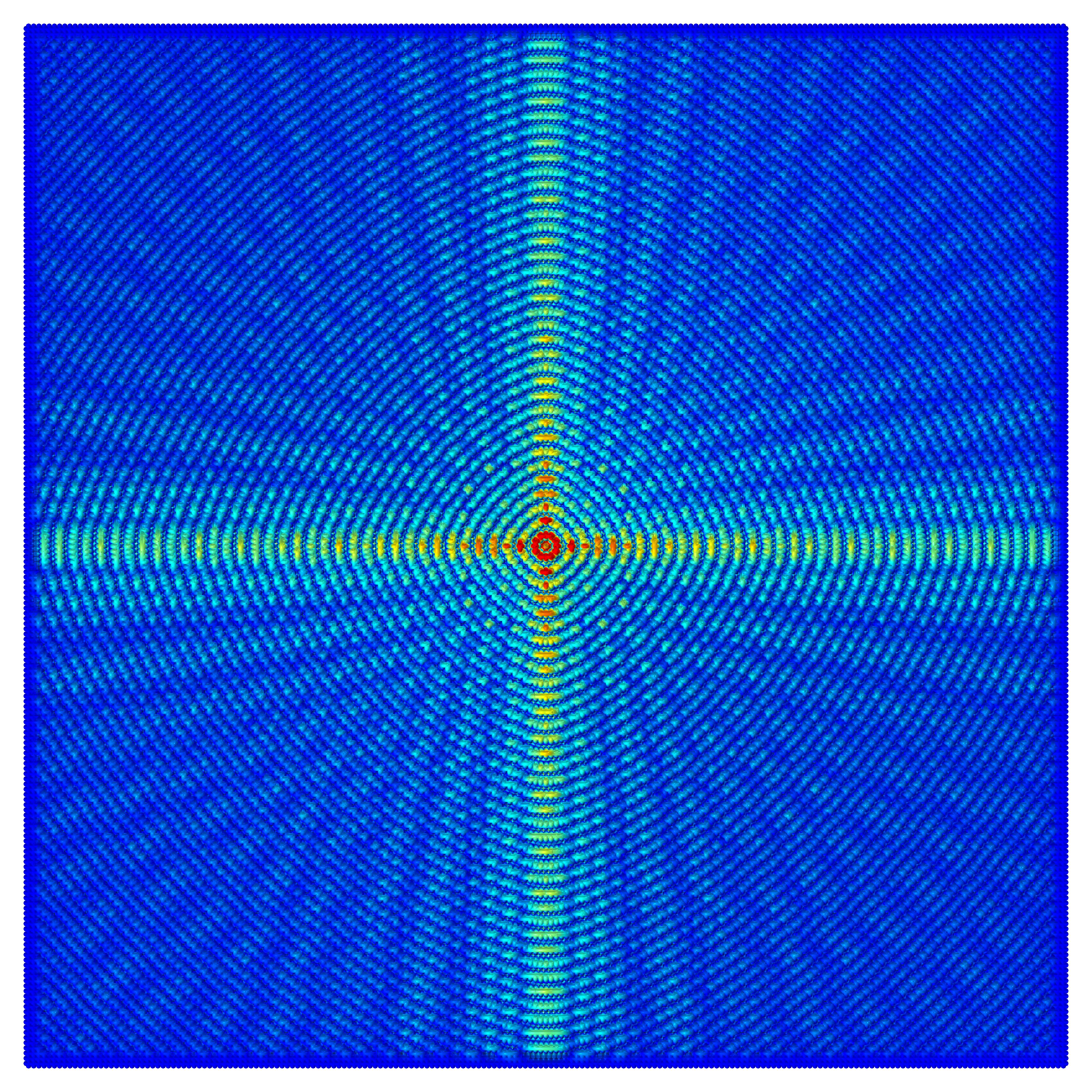}};
\node(b) at (a.south)  [
    anchor=center,
    xshift=0mm,
    yshift=-2mm
    ]{\begin{tikzpicture}
\pgfplotscolorbardrawstandalone[ 
    colormap/jet,
    colorbar horizontal,
    point meta min=0,
    point meta max=0.01,
    colorbar style={
        width=\rwa,
        height=0.2cm,
        xtick={0, 0.01},
        xticklabels={\small{0} },
        }]
\end{tikzpicture}
};
\node at (a.south east) [
    anchor=center,
    xshift=-7mm,
    yshift=-3mm
    ]{\small{0.01}};
        \end{tikzpicture}   
    \end{subfigure}%
    \hfill
    \begin{subfigure}[t]{0.32\textwidth}
        \centering
        \caption{$\omega=5$, $\vec{f} = [0,-1,0]^{\mathrm{T}}$}
        \label{FigCOMcombinedforceSR}
                \begin{tikzpicture}
\node(a){\includegraphics[width=\textwidth]{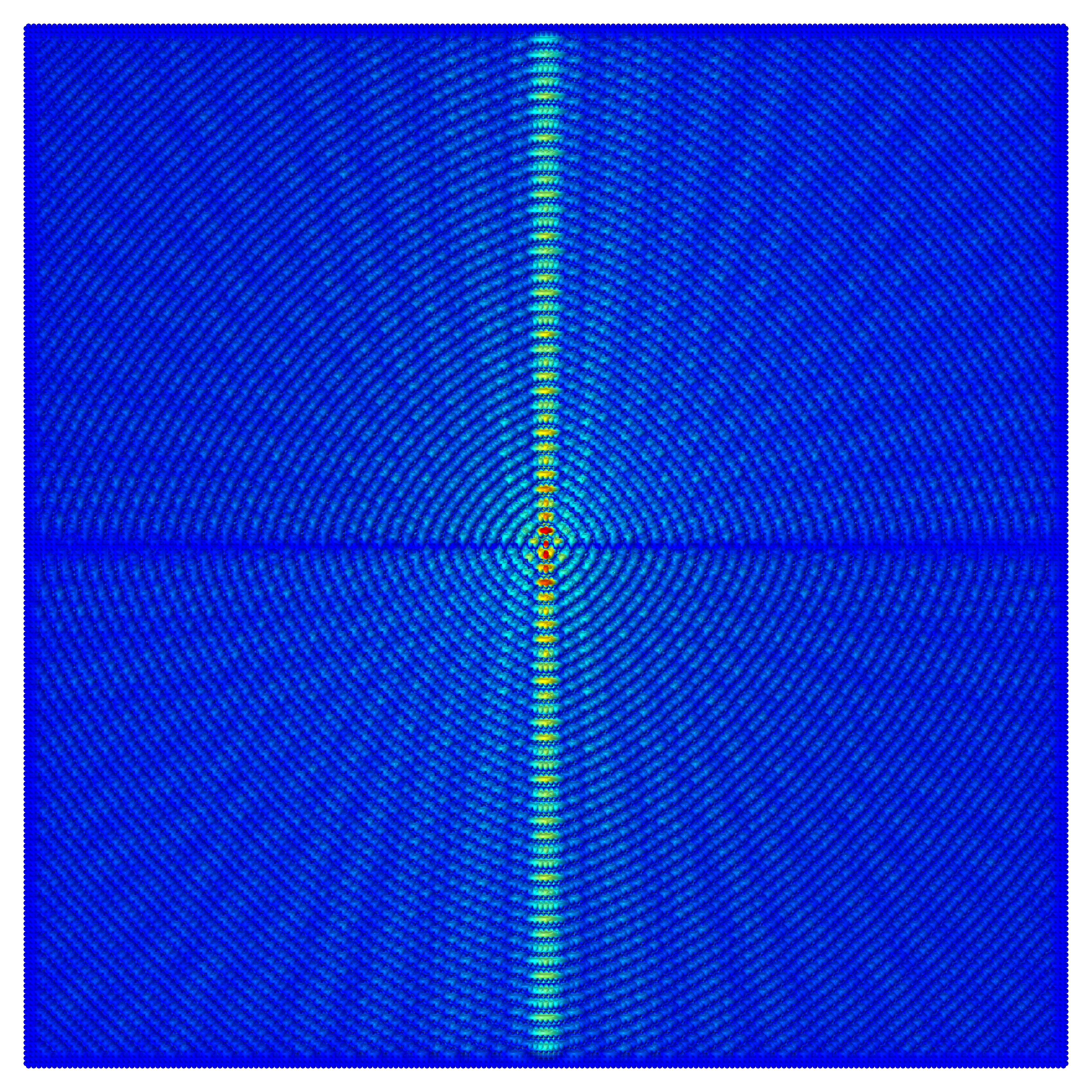}};
\node(b) at (a.south)  [
    anchor=center,
    xshift=0mm,
    yshift=-2mm
    ]{\begin{tikzpicture}
\pgfplotscolorbardrawstandalone[ 
    colormap/jet,
    colorbar horizontal,
    point meta min=0,
    point meta max=0.01,
    colorbar style={
        width=\rwa,
        height=0.2cm,
        xtick={0, 0.01},
        xticklabels={\small{0} },
        }]
\end{tikzpicture}
};
\node at (a.south east) [
    anchor=center,
    xshift=-8mm,
    yshift=-3mm
    ]{\small{0.025}};
        \end{tikzpicture}   
    \end{subfigure}
    \hfill
    \begin{subfigure}[t]{0.32\textwidth}
        \centering
        \caption{$\omega=5$, $\vec{f} = [-1,-1,0]^{\mathrm{T}}$}
      \label{FigCOMcombinedforceVSR}
        \begin{tikzpicture}
      \node(a){\includegraphics[width=\textwidth]{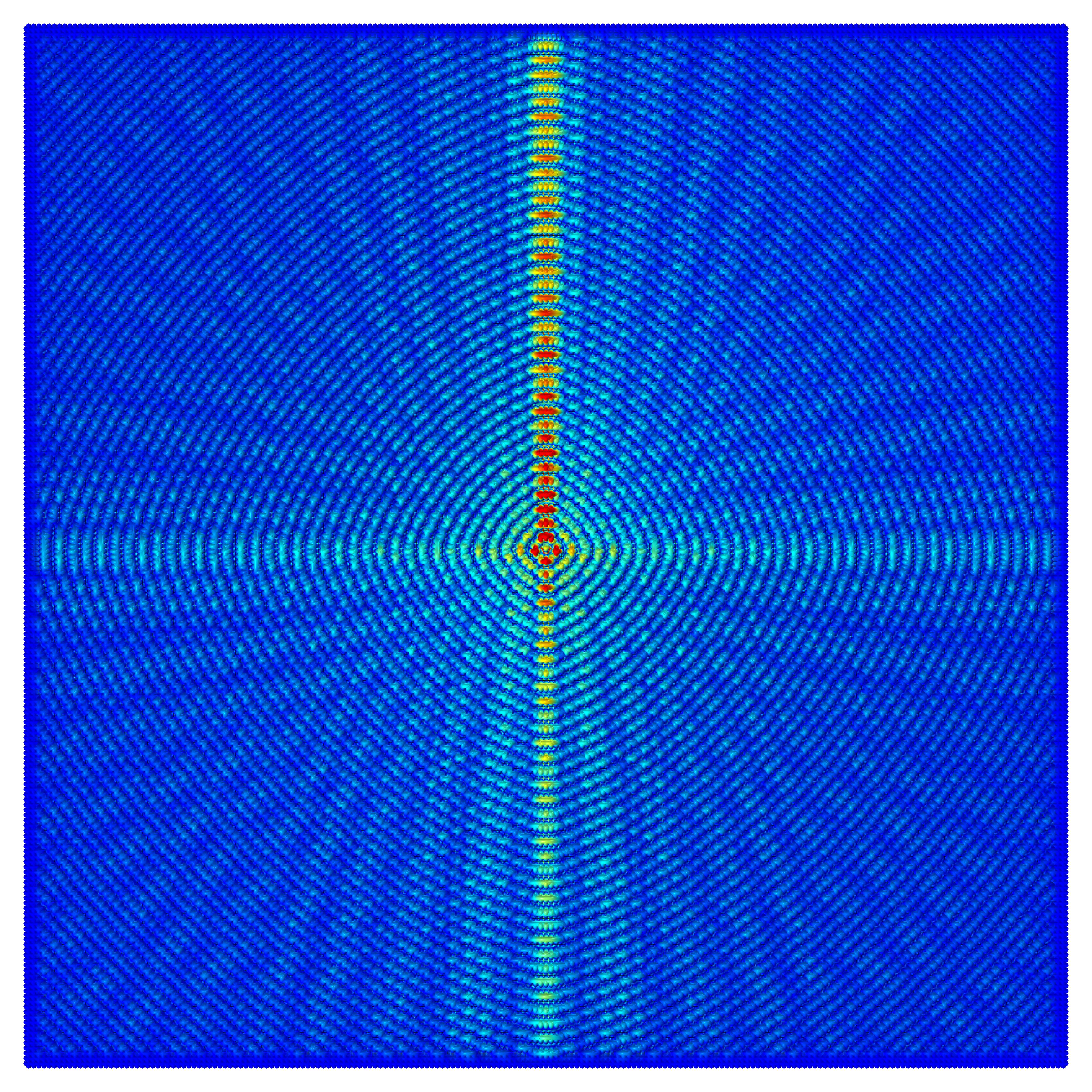}};
      \node(b) at (a.south)  [
    anchor=center,
    xshift=0mm,
    yshift=-2mm
    ]{\begin{tikzpicture}
\pgfplotscolorbardrawstandalone[ 
    colormap/jet,
    colorbar horizontal,
    point meta min=0,
    point meta max=0.01,
    colorbar style={
        width=\rwa,
        height=0.2cm,
        xtick={0, 0.01},
        xticklabels={\small{0} },
        }]
\end{tikzpicture}
};
\node at (a.south east) [
    anchor=center,
    xshift=-7mm,
    yshift=-3mm
    ]{\small{0.02}};
    \end{tikzpicture}
    \end{subfigure}
    \hfill
        \vspace{1.5em}
     \begin{subfigure}[t]{0.32\textwidth}
        \centering
        \caption{$\omega=8$, $\vec{f} = [-1,0,0]^{\mathrm{T}}$}
        \label{COMandSLOWw8Pic}
         \begin{tikzpicture}
\node(a){\includegraphics[width=\textwidth]{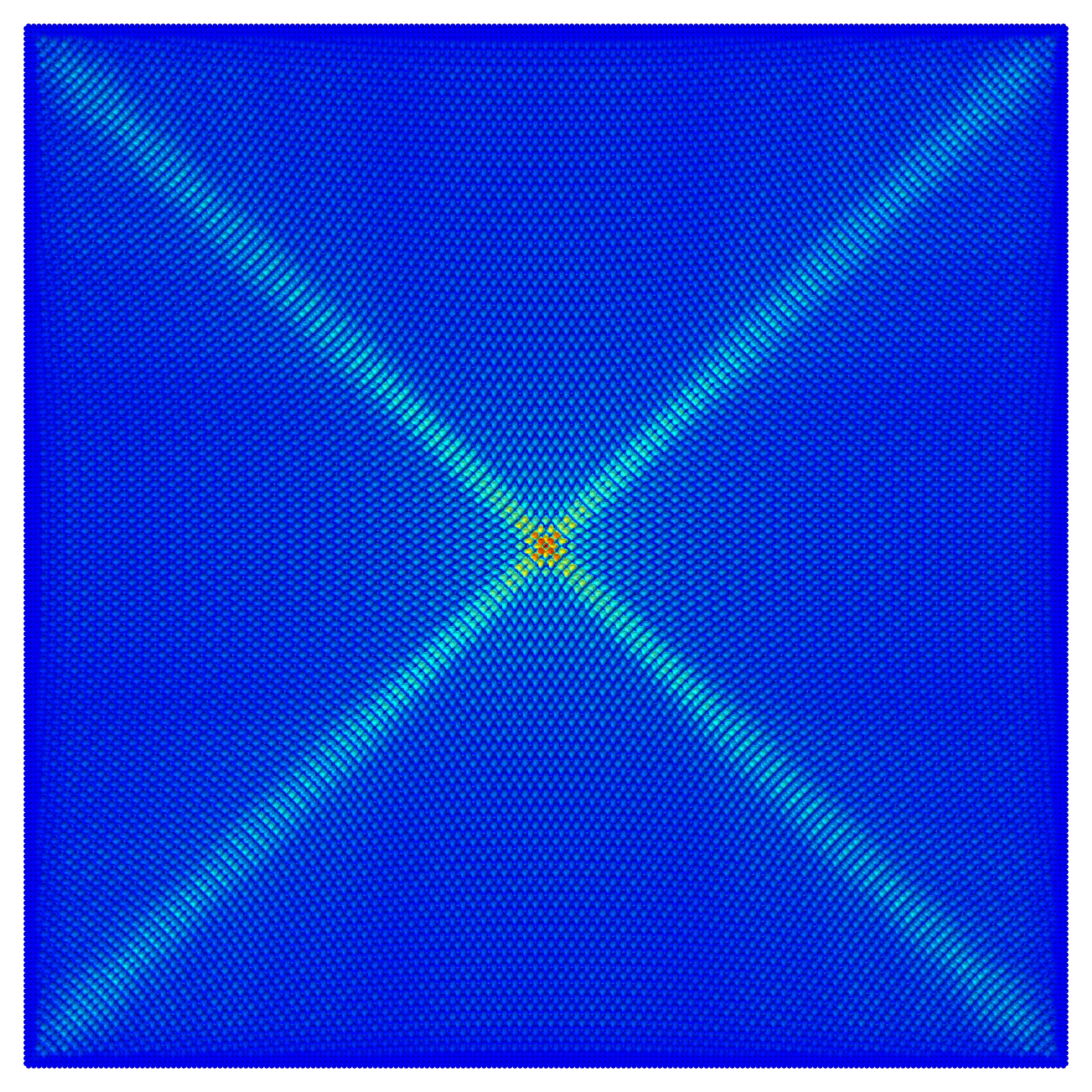}};
\node(b) at (a.south)  [
    anchor=center,
    xshift=0mm,
    yshift=-2mm
    ]{\begin{tikzpicture}
\pgfplotscolorbardrawstandalone[ 
    colormap/jet,
    colorbar horizontal,
    point meta min=0,
    point meta max=0.01,
    colorbar style={
        width=\rwa,
        height=0.2cm,
        xtick={0, 0.01},
        xticklabels={\small{0} },
        }]
\end{tikzpicture}
};
\node at (a.south east) [
    anchor=center,
    xshift=-8mm,
    yshift=-3mm
    ]{\small{0.008}};
        \end{tikzpicture}
    \end{subfigure}%
    \hfill
    \begin{subfigure}[t]{0.32\textwidth}
        \centering
        \caption{$\omega=8$, $\vec{f} = [-1,-1,-1]^{\mathrm{T}}$}
      \label{FigCOMcombinedforceVDR}
    \begin{tikzpicture}
\node(a){\includegraphics[width=\textwidth]{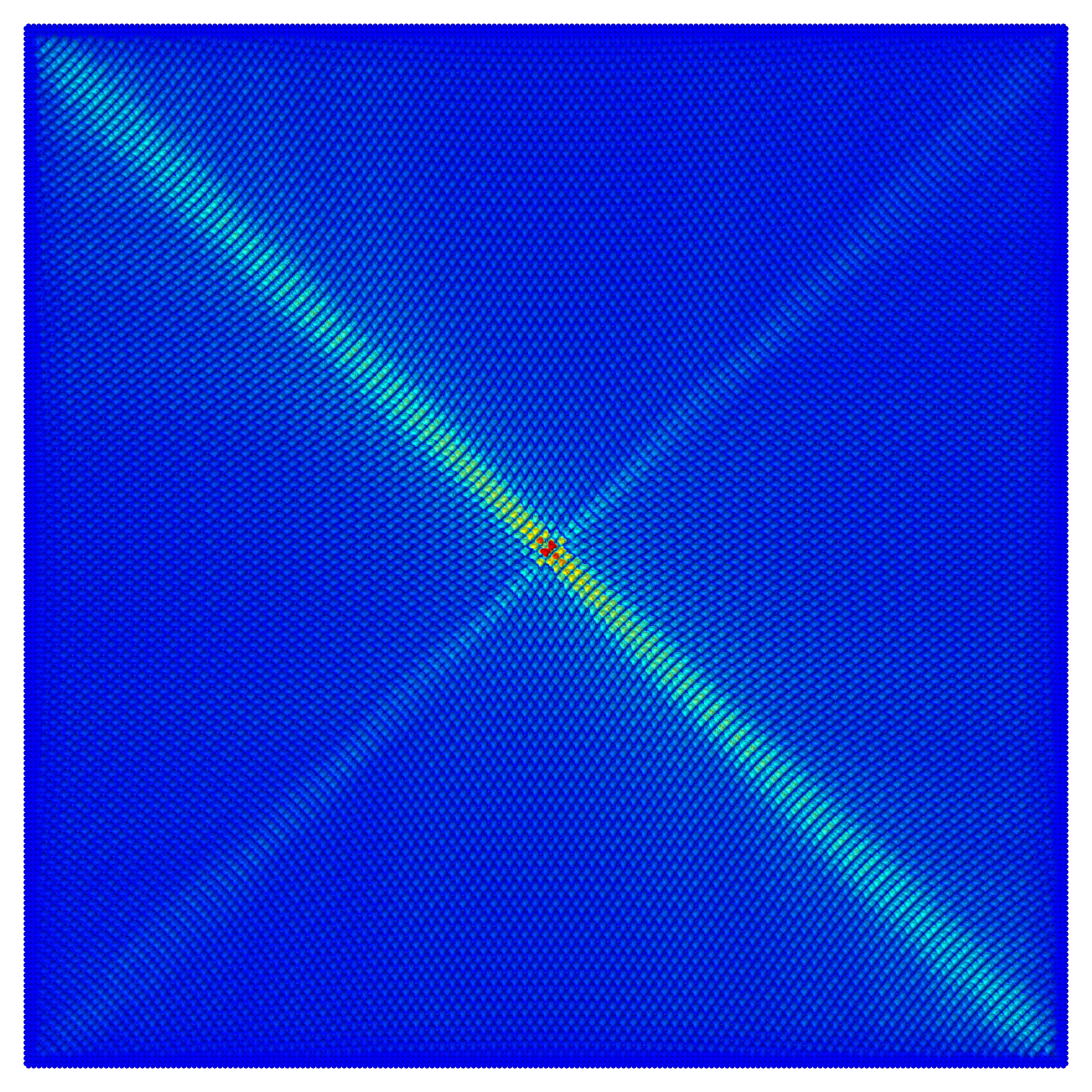}};
\node(b) at (a.south)  [
    anchor=center,
    xshift=0mm,
    yshift=-2mm
    ]{\begin{tikzpicture}
\pgfplotscolorbardrawstandalone[ 
    colormap/jet,
    colorbar horizontal,
    point meta min=0,
    point meta max=0.01,
    colorbar style={
        width=\rwa,
        height=0.2cm,
        xtick={0, 0.01},
        xticklabels={\small{0} },
        }]
\end{tikzpicture}
};
\node at (a.south east) [
    anchor=center,
    xshift=-8mm,
    yshift=-3mm
    ]{\small{0.02}};
        \end{tikzpicture}
    \end{subfigure}
    \hfill
    \begin{subfigure}[t]{0.31\textwidth}
        \centering
        \caption{Slowness contours} \label{SLOWw8and5}
        \begin{tikzpicture}
        \node[] at (0,0.2) {\includegraphics[width=\textwidth]{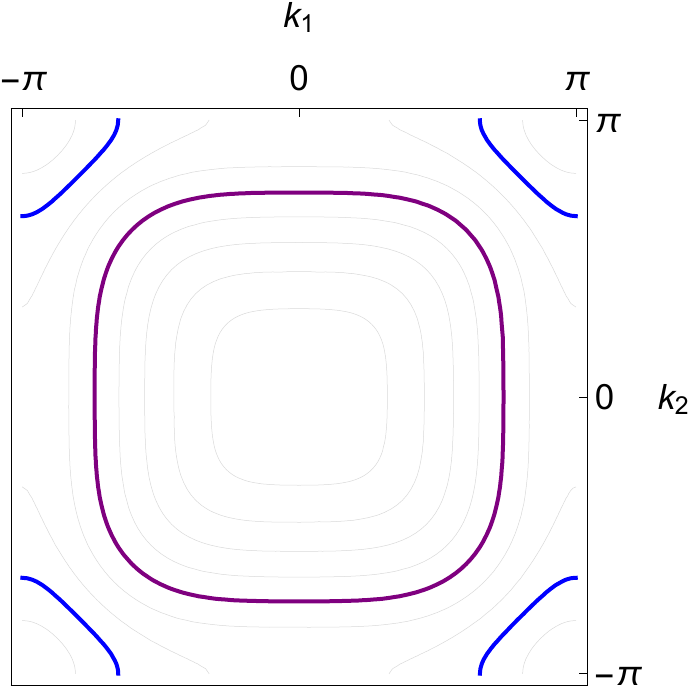}};
        \node[] at (-0.2,1.45) {\small $\omega=8$};
        \draw[<-, >=latex] (-1.5,1.4) -- (-0.8,1.4);
        \node[] at (0.4,-0.4)  {\small $\omega=5$};
        \draw[->, >=latex] (0.5,-0.5) -- (0.9,-0.9);
        \end{tikzpicture}
    \end{subfigure}
    \caption{The finite element model, with $\mu= 0.01$ and $C=0.1$, subject to the following forces and moments at $\vec{0}$ for the chosen frequency: (a) forcing $\vec{f} = [-1,0,0]^{\mathrm{T}}$ of frequency $\omega=5$; (b) forcing $\vec{f} = [0,-1,0]^{\mathrm{T}}$ of frequency $\omega=5$; (c) forcing $\vec{f} = [-1,-1,0]^{\mathrm{T}}$ of frequency $\omega=5$; (d) forcing $\vec{f} = [-1,0,0]^{\mathrm{T}}$ of frequency $\omega=8$; (e) forcing $\vec{f} = [-1,-1,-1]^{\mathrm{T}}$ of frequency $\omega=8$; and finally (f) the corresponding slowness contours for the lower dispersion surface of Figure~\ref{figlatticedispwithbandgap}.}
    \label{COMandSLOW}
\end{figure}
Figure~\ref{COMandSLOW} illustrates the use of the forcing vector to control the propagation of waves through the lattice for pass band frequencies.
In particular, we demonstrate that either uni-directional or asymmetric waves can be achieved by choosing the forcing vector appropriately. 
Firstly, in Figure~\ref{COMandSLOWw5Pic} we choose $\vec{f}=[-1,0,0]^{\mathrm{T}}$ corresponding to an out-of-plane force of frequency $\omega=5$. 
The corresponding slowness contour is provided in Figure~\ref{SLOWw8and5}; as expected, the principle directions of anisotropy coincide with the normals to the slowness contours, which themselves define the directions of maximal group velocity.
We note the stark contrast between Figures~\ref{COMandSLOWw5Pic} and~\ref{FigCOMcombinedforceSR} and emphasise that the only difference between the two figures is in the choice of forcing; the frequency of excitation is identical in both figures.
In particular, Figure~\ref{FigCOMcombinedforceSR} corresponds to $\vec{f}=[0,-1,0]^{\mathrm{T}}$, which is associated with the application of a clockwise moment about the $x$-axis.
The symmetric slowness contour indicates that a wave should propagate along both of the principle axes of the lattice equally, however when the lattice is forced by a moment about the $x$-axis, zero flexural displacement is induced in the $x$-axis, and a uni-directional waveform is induced which propagates in the $y$-direction only. 
Likewise, a uni-directional wave along the $x$-axis can be induced using the forcing vector $\vec{f}=[0,0,-1]^{\mathrm{T}}$.
In this way, the forcing vector can be used to select the direction of propagation, similar to the mechanism identified in~\cite{DynamicAnisotropy} for in-plane elastic systems.

The forcing vector also allows the application of combined forces and moments, producing modes with rare asymmetric anisotropy which can be used for applications in wave control and energy harvesting.
In Figure~\ref{FigCOMcombinedforceVSR}, the lattice was subjected to out-of-plane forcing with a simultaneous moment applied about the $x$-axis, such that $\vec{f}=[-1,-1,0]^{\mathrm{T}}$, at a frequency of $\omega=5$. 
The resulting wave propagates along the principle axes as expected, however with remarkably higher amplitude of displacement along the positive $y$-axis.
The combination of the applied translational force and rotational moment causes the $Y^{+}$ beam to displace more than the $Y^{-}$ beam, which induces the asymmetry.
Such non-reciprocity is rarely observed and usually requires PT-symmetry breaking but can sometimes be induced by creating asymmetric eigenmodes as in, for example,~\cite{NievesNon-ReciprocGyroLat}.

It is of note that while these effects have been demonstrated for the frequency $\omega=5$, these displacements can be achieved in a broad frequency regime. 
As can be seen in Figure~\ref{SLOWw8and5}, there is an extended interval of frequencies for which the slowness contours all have the same quasi-rectalinear shape. 
The waveforms at these frequencies all display similar localisation as Figure~\ref{COMandSLOWw5Pic}, and as such the same anisotropy can be induced through the choice of forcing vector; we further show in \S~\ref{Sec2DRotAdnTorsion} that the associated shape of the slowness contours is stable over a wide range of values of $\mu$ and $C$.
This is especially advantageous for the implementation of these effects in practical devices; in previous studies, generating uni-directional waves has required exact values of the material parameters and forcing frequency~\cite{FreeAndForced} and also has been associated with so-called `parabolic metamaterials' in narrow frequency regimes around Dirac cones~\cite{ParabolicMetamaterials}.
In contrast we have shown that, by fully taking into account the flexural and torsional interactions present in the lattice, it is possible to achieve similar effects without requiring highly precise tuning of material parameters.
This is particularly important for the fabrication of such devices where a degree of tolerance in the manufacturing process is required.

In Figure~\ref{COMandSLOW}, we also demonstrate that it is possible to control the dynamic anisotropy and shape of the localised waveforms propagating through lattice by altering the forcing frequency.
In Figure~\ref{COMandSLOWw8Pic}, we subject the lattice to an out-of-plane force of frequency $\omega=8$, producing a wave that travels equally along the diagonals of the lattice consistent with the principal directions predicted from the slowness contours in Figure~\ref{SLOWw8and5}. 
Comparing Figures~\ref{COMandSLOWw5Pic} and~\ref{COMandSLOWw8Pic}, it is observed that the principle directions of the two localised waveforms are rotated by $\pi/4$; the only difference between these two figures is the chosen forcing frequency. 

A further example, illustrating non-reciprocity, is shown in Figure~\ref{FigCOMcombinedforceVDR}, which gives the displacement under $\vec{f} = [-1,-1,-1]^{\mathrm{T}}$, inducing simultaneous out-of-plane forcing and double moments applied clockwise around both the $x$- and $y$-axes, at frequency $\omega=8$. 
While there is a small displacement propagating at $\pi/4$ to the principle axes, this is dominated by a much larger displacement propagating along the $3\pi/4$ line.
We also remark that the displacement along $3\pi/4$ is larger than along $7\pi/4$; although this is difficult to discern from Figure~\ref{FigCOMcombinedforceVDR}, it can be verified by direct evaluation of equation~\eqref{eq2DinverseFT}.

\subsection{The effects of altering the rotational inertia and the torsional stiffness}\label{Sec2DRotAdnTorsion}

We have shown above that the lattice provides significant control over the propagation of waves from the variety of forcing options alone, with rare examples of anisotropy.  
It is now shown that the dependence of the dispersion equation~\eqref{eq2Ddispersioneq} on the rotational inertia and torsional stiffness, not only provides extensive control over the location (and existence) of band gaps for the lattice, but also the shape of the dispersion surfaces and in turn the preferential direction of wave propagation. 
\begin{figure}[ht]
    \centering
    \begin{subfigure}[t]{0.45\textwidth}
        \centering
        \caption{}
        \includegraphics[width=\textwidth]{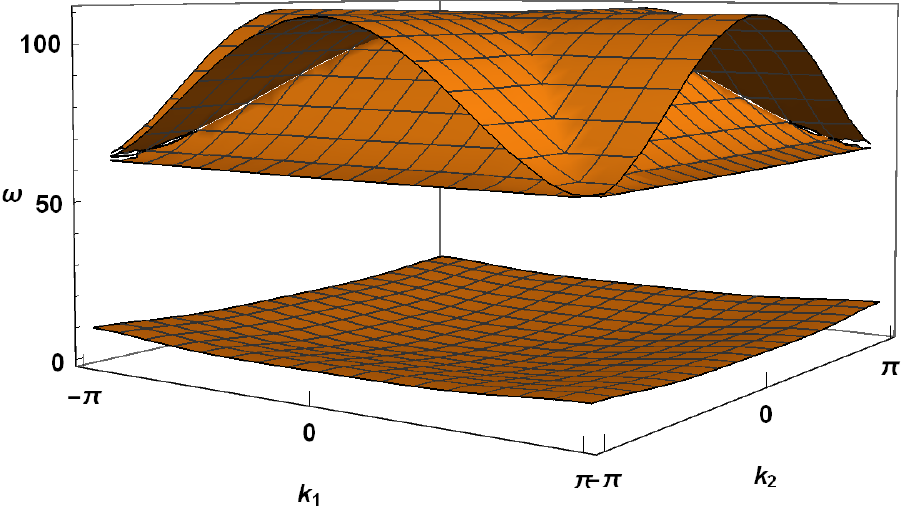}
        \label{DispdiagMu0,001c0,01}
    \end{subfigure}%
    \hfill
    \begin{subfigure}[t]{0.45\textwidth} 
        \centering
        \caption{}
        \includegraphics[width=\textwidth]{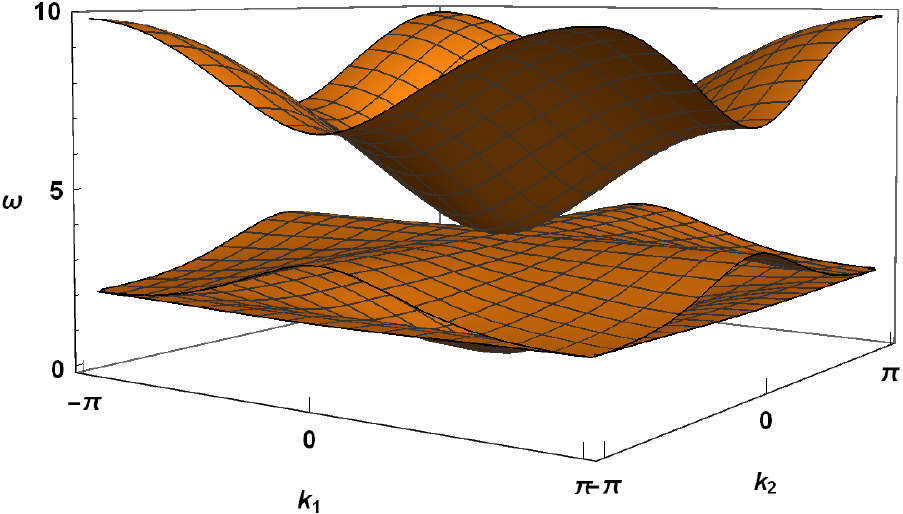}
       \label{DispdiagMu1c0,1}
    \end{subfigure}
    \hfill
    \begin{subfigure}[t]{0.45\textwidth}
        \centering
        \caption{}
        \includegraphics[width=\textwidth]{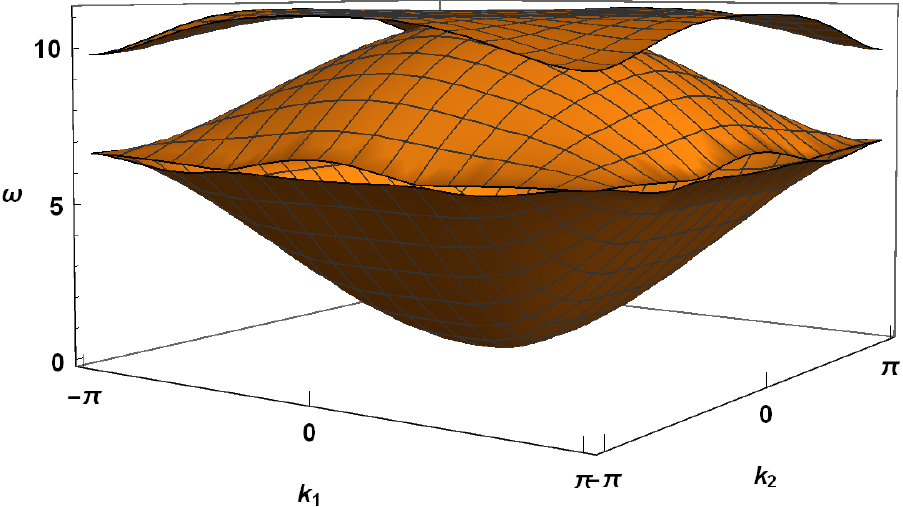}
        \label{DispdiagMu0,1c0,1}
    \end{subfigure}
    \hfill
    \begin{subfigure}[t]{0.45\textwidth}
        \centering
        \caption{}
        \includegraphics[width=\textwidth]{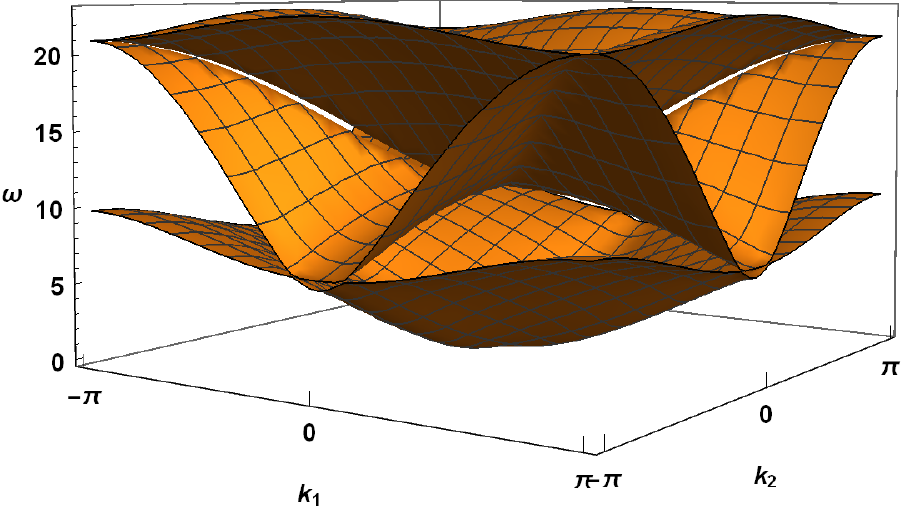}
        \label{DispdiagMu0,1c10}
    \end{subfigure}
    \caption{The dispersion diagrams for square lattices of Euler--Bernoulli beams with different values of rotational inertia $\mu$ and torsional stiffness $C$, where (a) $\mu=0.001$ and $C=0.01$ (b) $\mu=1$ and $C=0.1$  (c) $\mu=0.1$ and $C=0.1$ (d) $\mu=0.1$ and $C=10$.}
    \label{Disp3Diags}
\end{figure}

The different dispersion diagrams can be roughly classified into three shapes. 
When $\mu$ is very small (around 0.01 or below) and the torsional stiffness is allowed to range from very small to very large values, the dispersion diagrams have the same overall configuration as Figure~\ref{DispdiagMu0,001c0,01}, most often with a finite band gap. 
Generally the boundary of the acoustic band (the lower dispersion surface) does not change, although the shape of the surface does alter, in turn producing slowness contours with very different shapes.

When $\mu$ is 1 or larger, we observe a conical shape which is more commonly associated with beam lattices, such as in Figure~\ref{DispdiagMu1c0,1}. we also see that the boundary of the semi-infinite band gap usually occurs as much lower frequencies.
For intermediate values of $\mu$, the dispersion diagram can have a shape similar to Figure~\ref{DispdiagMu0,001c0,01}, although a finite band gap is seen less often, as is the case in Figure~\ref{DispdiagMu0,1c10}. Alternatively, we can see an entirely different shape, such as Figure~\ref{DispdiagMu0,1c0,1}, depending on the torsional stiffness.
Note that Figures~\ref{DispdiagMu0,1c0,1} and~\ref{DispdiagMu0,1c10} were produced using the same value of $\mu$, but different $C$.
This shows how amenable the dispersive properties of the lattice are to manipulation, indeed altering the values of $\mu$ and $C$ allows us to tailor the dispersion surfaces to desired choices of band gaps and slowness contours. 

It is noticed that, regardless of the size of the rotational inertia and torsional stiffness, for $k_{1}= k_{2} = \pi$, the solution to the dispersion equation is always $\omega = 4\sqrt{6}$; that is, $\sigma(4\sqrt{6}, \pi, \pi) =0$ for all values of $\mu$ and $C$. This $\omega = 4\sqrt{6}$ often forms the boundary to a band gap (see, Figures~\ref{figlatticedispwithbandgap},~\ref{DispdiagMu0,001c0,01} and~\ref{DispdiagMu1c0,1}), but not always (see, Figures~\ref{DispdiagMu0,1c0,1} and~\ref{DispdiagMu0,1c10}). 
Two other notable solutions to the dispersion equation are $\sigma(0, 0, 0) =0$, and $\sigma(4\sqrt{3}, 0, \pi) = \sigma(4\sqrt{3}, \pi, 0) =0$ for all values of $\mu$ and $C$. 
In particular, the frequency $\omega = 4\sqrt{3}$ at the edge of the Brillouin zone provides a clear connection to the one-dimensional case, (cf. \S~\ref{secdispeq}).


\section{A chain of Euler--Bernoulli beams}\label{BeamChain}

In this section, we construct a new class of Green's functions for an infinite one-dimensional chain of thin Euler--Bernoulli beams with connecting junctions that possess both mass and rotational inertia.
Unlike the square lattice studied in \S~\ref{sec2DLattice}, there are no torsional interactions to account for in the 1D chain. 
This simplification, along with having only one Fourier parameter means that the Green's functions can be evaluated in closed form, as will be shown. 
In a right-handed coordinate system, we construct the chain along the $y$-axis, and define the translational motion along the $z$-axis, as shown in Figure~\ref{figEBCdiagram}. 
While there are no torsional interactions, each junction in the chain has two degrees of freedom, corresponding to the translational displacement $w$ and rotational displacement (about the $x$-axis) $\theta_{x}$. 

\newlength{\rwd}
\setlength{\rwd}{0.8\textwidth}%
\begin{figure}[ht]
\centering
\begin{tikzpicture}
\node[] at (0,0) {\includegraphics[width=\rwd]{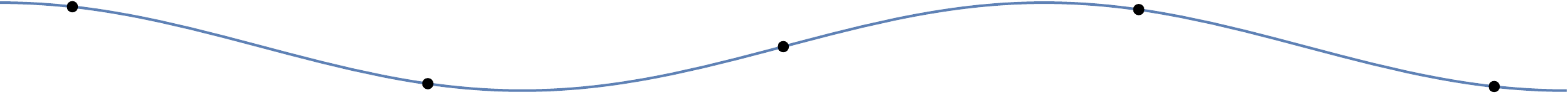}};
\draw[->, >=angle 60] (-{0.5*\rwd},0) -- ({0.5*\rwd},0);
\draw[->, >=angle 60] (0,-0.6) -- (0, 1.3);
\draw[->, >=stealth, line width=1] (0,0) -- (0,1);
\node[] at (5.7,0)  {\small $y$};
\node[] at (0,1.4)  {\small $z$};
\node[] at (-0.25,0.75)  { $w$};
\node[] at (0.47, 0.4)  { $\theta_x$};
\tikzset{
    partial ellipse/.style args={#1:#2:#3}{
        insert path={+ (#1:#3) arc (#1:#2:#3)}    }}
\draw[->, >=stealth, line width=1] (0,0) [partial ellipse=65:405:0.35 and 0.35]; 
\end{tikzpicture}
\caption{The flexural displacement of an Euler--Bernoulli beam chain, lying along the $y$-axis, with rotations measured anticlockwise about the $x$-axis (out of the page).}
\label{figEBCdiagram}
\end{figure}
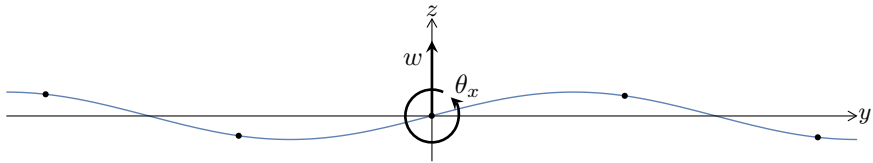

As before, we introduce the generalised displacement vector $\vec{u}_{n}=\left[w(n),\theta_{x} (n)\right]^{\mathrm{T}}$ of the $n^\mathrm{th}$ junction to describe the translational and rotational displacements of the nodes, which are coupled through the flexural deformations of the beams.
The mass of the nodes, flexural rigidity and length of the beams are once again chosen as natural units.
To construct the equation of motion for the $n^\mathrm{th}$ unit cell, we evaluate the forces acting on the node at $y=n$ from the beam on $n\leq y \leq n+1$ which we denote $Y^{+}$, using the same method as previously. 

The flexural deformation of the Euler--Bernoulli beams, $w (y)$ is governed by the same fourth order differential equation used in \S~\ref{sec2DLattice}, equation~\eqref{2Deq4thorderbeam}.
As before, $w(y)$ is a cubic polynomial in $y$, which is found using boundary conditions for the ends of the $Y^{+}$ beam,
\begin{multicols}{2}
\noindent
\begin{eqnarray*}
   w(n) &=& W_{0}\\
   w_{,x}(n) &=& \Theta_{0}
\end{eqnarray*}
\begin{eqnarray*}
   w(n+1) &=& W_{1}\\
   w_{,x}(n+1) &=& \Theta_{1};
   \label{constants}
\end{eqnarray*}
\end{multicols}
\noindent 
where $W_{i}$ are the translational displacements and $\Theta_{i}$ are the angles of rotation of the junctions. 
The positive flexural rotations of each junction are defined anticlockwise about the $x$-axis, shown in Figure~\ref{figEBCdiagram}. 
We use the same equations for the shear force $F(y)$ and bending moment from the flexural rotations $M_{flex}(y)$ as in \S~\ref{sec2DLattice}, equation \eqref{2DeqForceMoments}.
We introduce the generalised forcing vector $\mathbb{F}_n = [ F(n), M_{flex}(n)]^\mathrm{T}$ and as such, the force and the bending moment at $y=n$ associated with the $Y^{+}$ beam can be expressed as
\begin{equation}
\mathbb{F}_{n} \big|_{Y^{+}} = A \vec{u}_{n} + B \vec{u}_{n+1}
\label{eqFMzerounitcell}
\end{equation}
where the matrices
\begin{equation}
    A = 
\begin{bmatrix} 
-12 & -6 \\
-6 & -4 
\end{bmatrix}
\quad
\text{and}
\quad
B=
\begin{bmatrix} 
12 & -6 \\
6 & -2 
\end{bmatrix}\nonumber
\end{equation}
encapsulate the flexural rigidity of the beams connecting the $n^\mathrm{th}$ node to the $(n+1)^\mathrm{th}$ node.

Considering Figure~\ref{figEBCdiagram}, it can be seen that the translational forces at the $y=n+1,n-1$ nodes have opposite sign, while the bending moments have same direction. 
Hence we use the matrix $R=\diag[-1,1]$, which describes a rotation by $\pi$ of the coordinate system about the $x$-axis.
Using this rotation, the forces exerted on the node at $y=n$ by the beam that lies on $n-1\leq y \leq n$ (which we denote $Y^{-}$), are
\begin{equation}
    \mathbb{F}_{n}\big|_{Y^{-}}  =  RAR^{-1}\vec{u}_{n} + RBR^{-1}\vec{u}_{n-1}.
\end{equation}
With Newton's second law, the time-harmonic equation of motion of the $n$\textsuperscript{th} node can be expressed
\begin{equation}
0 = \left[\omega^{2} \mathsf{M} + A + R A R^{-1} \right]\vec{u}_{n}+ B\,\vec{u}_{n+1} + R B R^{-1}\vec{u}_{n-1},
\label{eqnthun}
\end{equation}
where $\mathsf{M}=\diag[1,\mu]$ is the matrix that describes the inertial properties of the chain.
The first component of $\mathsf{M}$ corresponds to the mass of the junction whilst the second component corresponds to the rotational inertia, denoted $\mu$. 
In the chosen system of natural units $\mu$ is actually the moment of inertia per unit mass of the nodes and so, for junctions between thin beams, $0<\mu\ll1$.
After applying the discrete Fourier transformation
\begin{equation}
\vec{u}^{F}(k) = \sum_{n\in\mathbb{Z}} \mathrm{e}^{-\I k n}\vec{u}_{n} \, ,
\label{1DeqFT}\nonumber
\end{equation}
to equation~\eqref{eqnthun}, we arrive at the equation of motion for the Euler--Bernoulli beam chain in reciprocal space,
\begin{equation}
0 = \left[\omega^{2} \mathsf{M} + A + R A R^{-1} + \mathrm{e}^{\I k}B + \mathrm{e}^{-\I k}R B R^{-1} \right] \vec{u}^{F}(k),
\label{BeamFT}
\end{equation}
where $k$ is the Fourier variable and $\vec{u}^{F}(k) = [W^{F}(k), \Theta^{F}(k)]^{\mathrm{T}}$ is the generalised displacement in reciprocal space, formed of the Fourier transformation of the real space displacements $w(n)$ and $\theta_{x}(n)$.


\subsection{The dispersion equation}\label{secdispeq}

The solvability condition of equation~\eqref{BeamFT} yields the dispersion equation, $\sigma(\omega,k) =0$, where
\begin{equation}
   \sigma(\omega,k) =  \mu\, \omega^{4} + ( 24 \mu \cos(k) - 4 \cos(k) - 24 \mu - 8)\,\omega^{2}+ 24 \cos (2k) - 96\cos(k) + 72 .
   \label{eqbeamchaindisp}
\end{equation}
Given that equation~\eqref{eqbeamchaindisp} is quadratic in $\omega^{2}$, we can find analytical representations $\omega=\omega(k)$ for the dispersion curves, such that $\sigma(\omega(k),k) = 0$.
Moreover, the exact values of the edges of the band gaps can be found easily.
The dependence of the dispersion equation on $\mu$ illustrates the importance of the rotational inertia; in particular the rotational inertia can be used as a parameter to control the location and width of the band gaps.

\begin{figure}[ht]
\centering
    \includegraphics[width=0.6\textwidth]{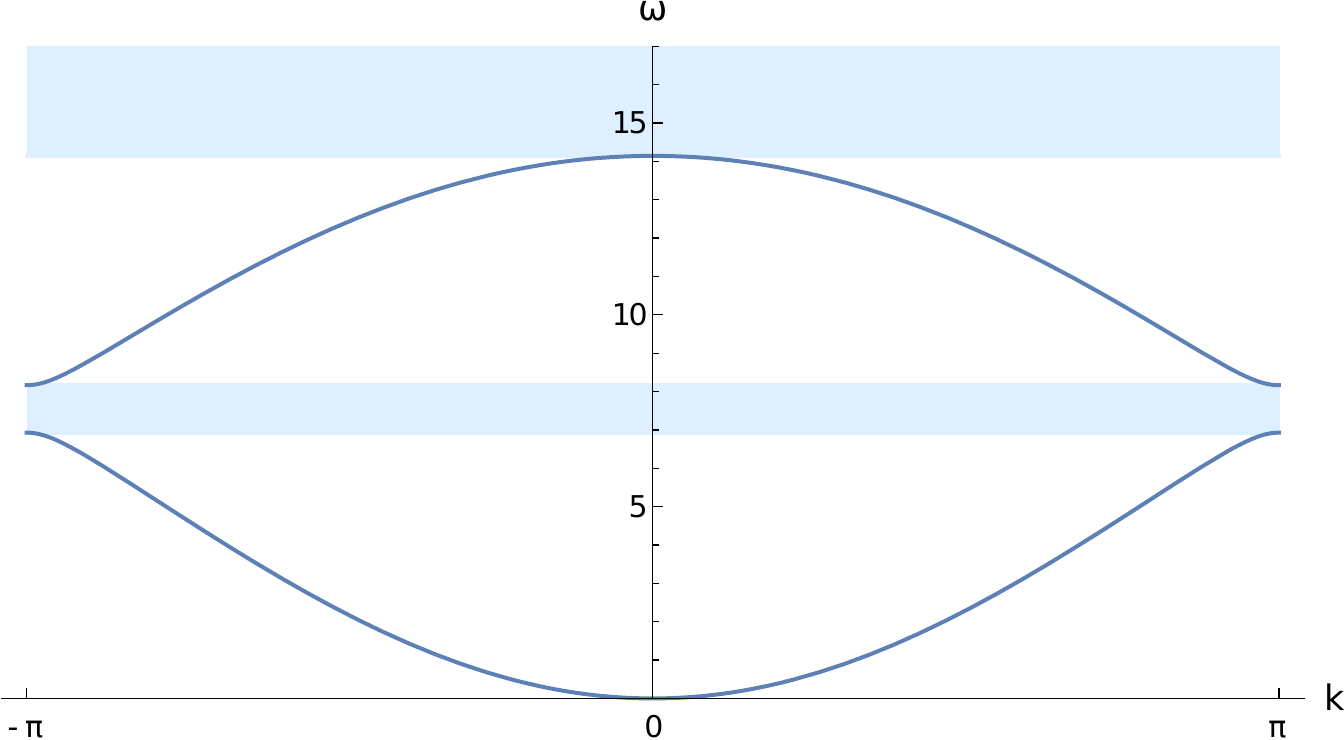}
    \begin{minipage}{\textwidth}
    \caption{Dispersion curves for the chain of Euler--Bernoulli beams, for the chosen value of $\mu = 0.06$. The band gaps have been shaded light blue.}
    \label{figbeamchaindisp}
    \end{minipage} 
\end{figure}

In Figure~\ref{figbeamchaindisp} we plot the dispersion diagram for a chain with $\mu = 0.06$, across the first Brillouin zone $k \in [-\pi,\pi]$. 
It is seen that, for $\mu=0.06$, the Euler--Bernoulli beam chain has two band gaps, one finite band gap for $4\sqrt{3}< \omega < 10 \sqrt{2/3}$, and the semi-infinite band gap associated with discrete systems, for $\omega > 10\sqrt{2}$.

\begin{figure}[ht]
\centering
\begin{minipage}{\textwidth}
    \centering
    \includegraphics[width=0.6\textwidth]{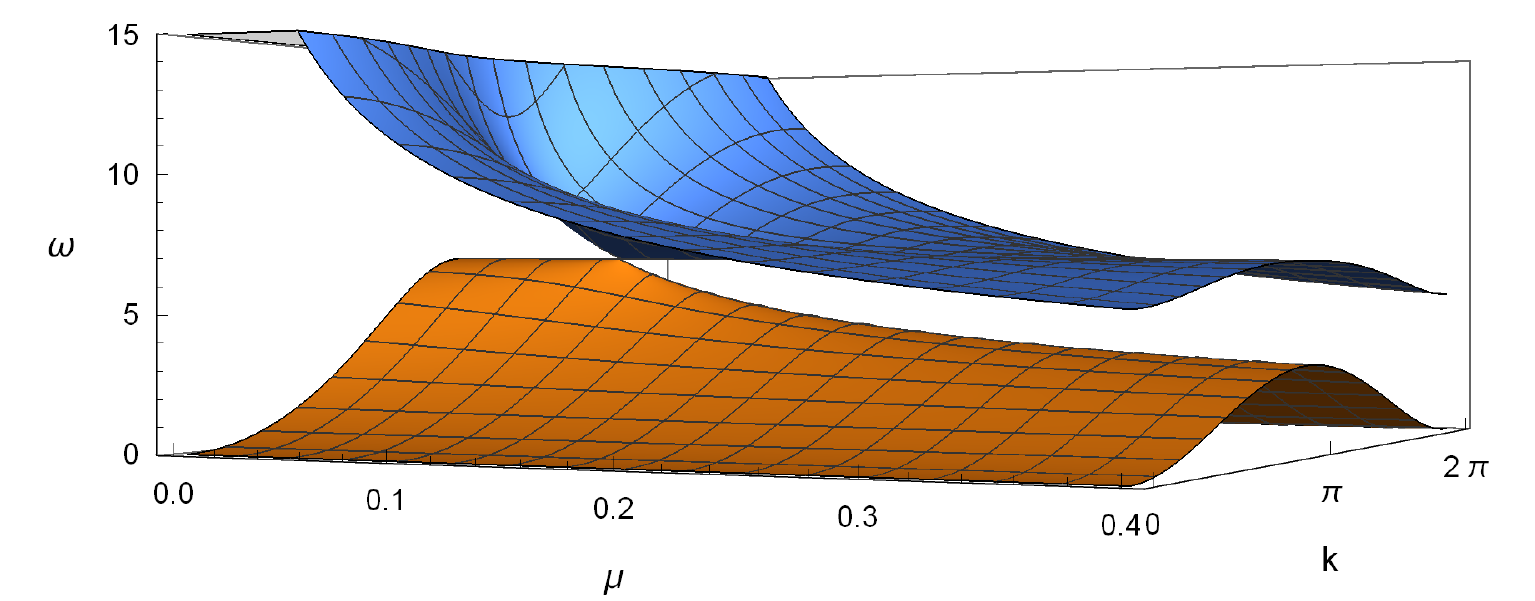}
     \caption{Planes have been formed from the dispersion curves for the Euler--Bernoulli beam chain across a range of $\mu$. The planes do not touch apart from for one degenerate value of $\mu = 1/12$ when $\omega = 4 \sqrt{3}$.}
     \label{Degenmu}
\end{minipage}
\end{figure} 
In Figure~\ref{Degenmu}, we explore the evolution of the dispersion curves with changing rotational inertia.
It is emphasised that Figure~\ref{Degenmu} does not show dispersion surfaces, rather planes have been formed by plotting the dispersion curves for a range $\mu$.
The planes are disjoint except for at one degenerate value of $\mu = 1/12$ when $\omega = 4 \sqrt{3}$.
For any $\mu \neq 1/12$, the chain always has two band gaps: a semi-infinite band gap for large $\omega$, and also a finite band gap with non-zero width. 
Even as $\mu \to \infty$, a finite band gap still exists, although it becomes infinitesimally thin.
Furthermore, regardless of the value of $\mu$, the dispersion equation always satisfies $\sigma(4\sqrt{3},\pi)=0$. 
This provides a connection to the square lattice (c.f. \S~\ref{sec2DLattice}) since $\sigma(4\sqrt{3}, 0, \pi) = \sigma(4\sqrt{3}, \pi,0) = 0$ and, in this regime, the 2D problem becomes quasi-one-dimensional.
When $\mu=1/12$ the semi-infinite band gap remains and, despite the fact there is no finite band gap, forcing the chain at this degeneracy nevertheless results in a localised mode, as discussed in \S~\ref{SECDegeneratemu}.
In the vicinity of $\mu=1/12$, the critical point $( \omega, k) = (4\sqrt{3}, \pi)$ is associated with the lower/upper boundaries of the finite band gap for $\mu\lessgtr 1/12$, and for $\mu\gg1/12$, $\omega = 4 \sqrt{3}$ forms the lower boundary of the semi-infinite band gap.


\subsection{The Green's function}\label{sec1DBGGF}

To construct the Green's function, we consider the application of force $\vec{f}\in\mathbb{C}^{2}$ to the central node of the lattice.
The forcing vector $\vec{f}=\left[f_{w},f_{\theta}\right]^{\mathrm{T}}$ has components corresponding to the translational force $f_{w}$ and rotational moment $f_{\theta}$, and we let $f_{w},f_{\theta}\in\left\{0,-1\right\}$ with the exclusion of $f_{w} = f_{\theta} = 0$ which equates to the absence of an applied force.
The Green's function in reciprocal space can then be expressed
\begin{equation}
\vec{u}^{F}(k) = \left[A + R A R^{-1} + \omega^{2} \mathsf{M} + \mathrm{e}^{\I k}B + \mathrm{e}^{-\I k}R B R^{-1} \right]^{-1} \vec{f}.
\label{eqmatrixoperator}
\end{equation}
For convenience, we separate the flexural displacement $W^{F}(k)$ and rotation $\Theta^{F}(k)$ components for evaluation. 
Explicitly, the flexural displacement of the chain in reciprocal space is 
\begin{equation}
    W^{F}(k)= \frac{ (\mu \omega^{2}-4 \cos(k) -8 )f_{w} + 12\I \sin(k)f_{\theta} }{\sigma(\omega,k)} \, .
    \label{eqWFKintegrand}
\end{equation}
The denominator of $W^{F}(k)$ coincides with the dispersion equation~\eqref{eqbeamchaindisp} as a result of the inverted matrix in equation~\eqref{eqmatrixoperator}.
The flexural displacement in direct space is obtained by applying the inverse Fourier transform to the spectral representation equation~\eqref{eqWFKintegrand}, as follows
\begin{equation}
    w(n) = \frac{1}{2\pi}\int\limits_{-\pi}^{\pi} W^{F}(k)\, \mathrm{e}^{\I kn} \dd k \, .
    \label{eqWXintegral}
\end{equation}
A significant difference between the chain of beams and the square lattice from \S~\ref{sec2DLattice} is the ability to evaluate the integral of the inverse Fourier transformation in closed form. 
To evaluate the integral we use the substitution $z = \mathrm{e}^{\I k}$ to map the line segment $[-\pi,\pi]$ to the unit circle in the complex plane. 
Following the substitution, we denote the displacement $W^{F}(z)$, and apply Cauchy's Residue Theorem by determining where the poles of the integrand, $z_{i}$, lie in respect to the unit circle and taking the residues as follows, 
\begin{equation}
    w(n) = \frac{-\I}{2\pi} \oint z^{n-1} W^{F}(z) \dd z    \,=\, \sum_{\abs{z_{i}} < 1} \Res [W^{F}(z), z_{i}] - \frac{1}{2}\sum_{\abs{z_{i}} = 1} \Res [W^{F}(z), z_{i}].
    \label{eq1DResidueeq}
\end{equation}
The numerator of equation~\eqref{eqWFKintegrand} is an entire function and, therefore, the poles of equation~\eqref{eqWXintegral} are the zeros of the dispersion equation~\eqref{eqbeamchaindisp} and, consequently, are frequency dependent.
Nevertheless, we can use the fact that the complex solutions of the dispersion equation only cross the unit circle when the frequency traverses the boundary of a pass band to construct closed form Green's functions for each frequency regime: finite band gap, semi-infinite band gap and pass band frequencies.


\subsection{Finite band gap}

\begin{figure}[H]
\centering
\begin{minipage}{\textwidth}
    \centering
    \includegraphics[width=0.85\textwidth]{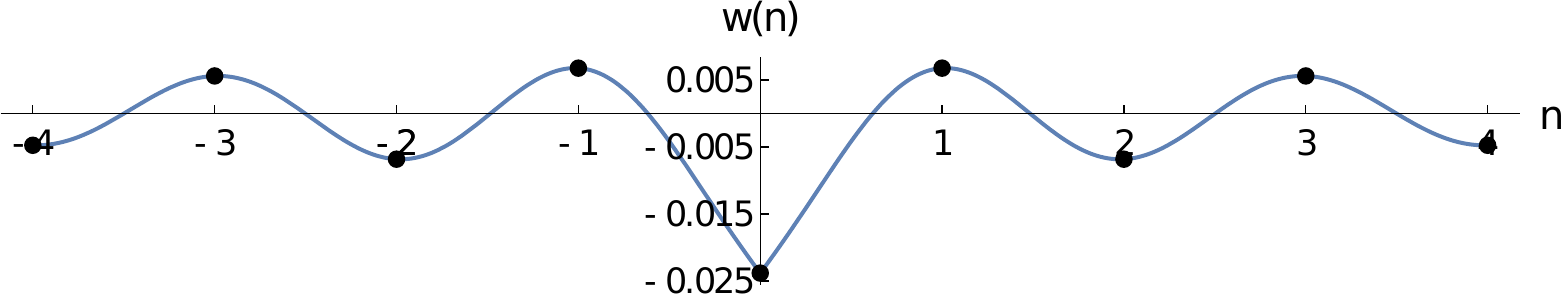}
    \caption{A plot of the flexural displacement Green's function $w(n)$ in the finite band gap of the Euler--Bernoulli beam chain subject to point forcing $\vec{f} = [-1,0]^{\mathrm{T}}$ at the origin, where $\mu=0.06$ and $\omega = 7.5$. }
     \label{vertmotionplot}
\end{minipage}
\end{figure}

As an example, we evaluate explicitly the flexural displacement Green's function in the finite band gap regime for translational forcing $\vec{f} = [-1,0]^{\mathrm{T}}$, using equation~\eqref{eq1DResidueeq}.
For frequencies in the finite band gap, 
\begin{equation}
   w(n) =  \frac{\rho_{+}\gamma_+^n
   + \rho_{-} \gamma_-^n}{2^{\left(3 n+\frac{1}{2}\right)} 3^{n}\psi\omega^2\left(\omega^{2} -48 \right)},
   \label{eqclosedform}
\end{equation}
where the following repeated factors have been introduced
\[
    \psi = \sqrt{144 + (1 - 24 \mu + 36 \mu^2) \omega^2}\, ,
    \quad
    \gamma_{\pm} = 24 + (1-6\mu)\omega^{2} \pm \psi \omega \mp\sqrt{2} \rho_{\pm}\,  \quad \text{and}
\]
\[
    \rho_{\pm} = \left(6 \mu  \omega - 2\omega \pm \psi\right)\sqrt{\left(36 \mu ^2-18 \mu +1\right) \omega ^4 \pm (\psi -6 \mu  \psi )\omega ^3 +(96-144 \mu ) \omega ^2 \pm 24 \psi  \omega } \,. 
\]
In Figure~\ref{vertmotionplot}, we plot $w(n)$ for the chosen values of $\mu=0.06$ and $\omega=7.5$, it can be seen that the waves decay exponentially away from the forcing point at the origin as expected.

\subsection{Rotational forcing}

Choosing the forcing vector $\vec{f}= [0,-1]^{\mathrm{T}}$ in equation~\eqref{eqmatrixoperator} has the effect of applying a clockwise moment about the $x$-axis, rather than the previous translational force. 
We then evaluate the real space Green's function in the same manner, using equation~\eqref{eq1DResidueeq}, to find the flexural displacement of the chain in response to the applied moment. 
This is plotted in Figure~\ref{EBCfiniteW9,8Rot}, using the same values for $\mu=0.06$ and $\omega=7.5$ as were used in Figure~\ref{vertmotionplot} for comparison.

\begin{figure}[ht]
\centering
\begin{minipage}{\textwidth}
    \centering
    \includegraphics[width=0.85\textwidth]{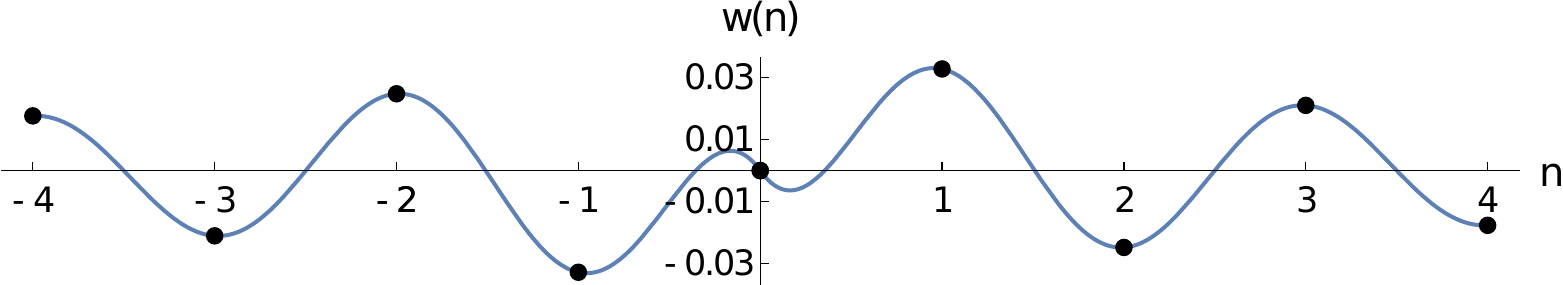}
    \caption{A plot of the flexural displacement Green's function $w(n)$ for $\mu=0.06$ and $\omega = 7.5$ within the finite band gap, induced by an applied moment about the $x$-axis $\vec{f} = [0,-1]^{\mathrm{T}}$, at $n=0$.}
     \label{EBCfiniteW9,8Rot}
\end{minipage}
\end{figure} 

Figure~\ref{EBCfiniteW9,8Rot} clearly demonstrates the twist induced in the chain from the rotational forcing. 
There is notably zero translational motion at the $n=0$ node, as expected for a purely rotational force, followed by a comparatively large displacement, anti-symmetric in each direction, which soon decays. 


\subsection{The semi-infinite band gap}

\begin{figure}[ht]
\begin{minipage}{\textwidth}
    \centering
    \includegraphics[width=0.85\textwidth]{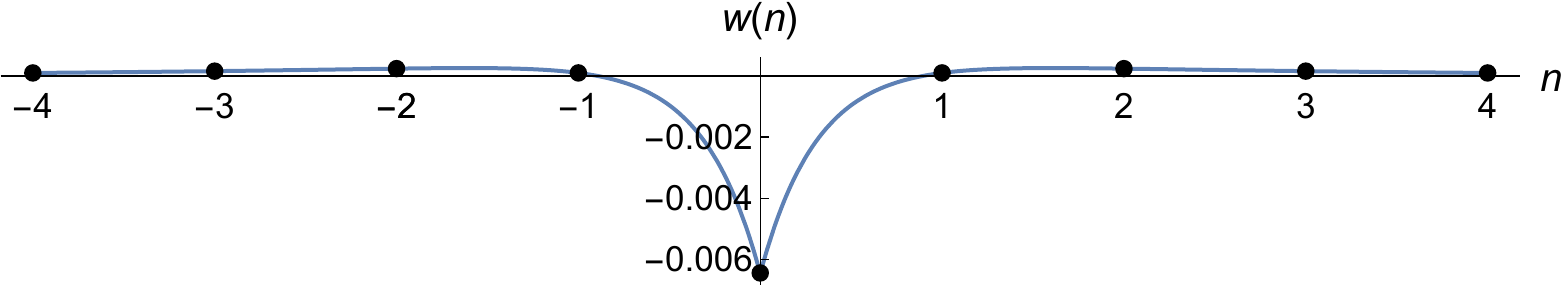}
    \caption{A plot of the flexural displacement Green's function $w(n)$ for $\mu=0.06$ and $\omega = 14.3$ in the semi-infinite band gap, due to point forcing $\vec{f} = [-1,0]^{\mathrm{T}}$ at $n=0$.}
     \label{1DseminfVDVF}
\end{minipage}
\end{figure}

\begin{figure}[ht]
\centering
\begin{minipage}{\textwidth}
    \centering
    \includegraphics[width=0.85\textwidth]{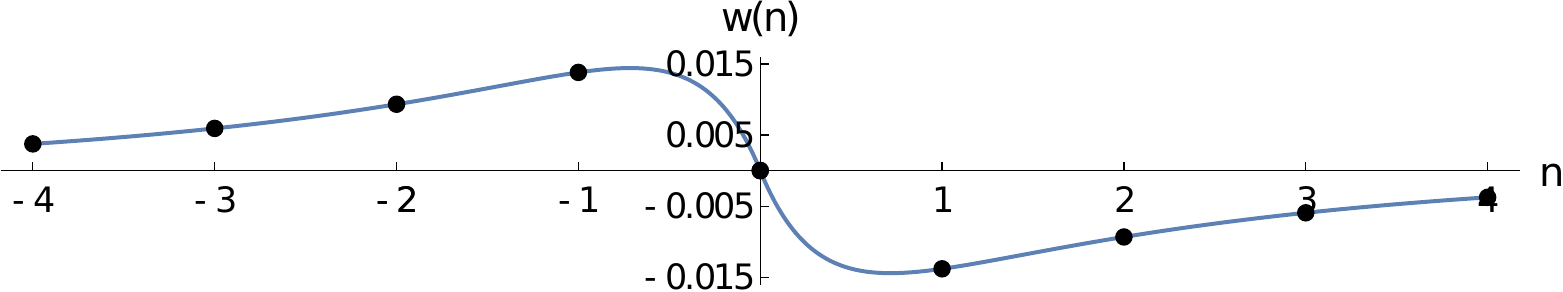}
    \caption{A plot of the flexural displacement Green's function $w(n)$ for $\mu=0.06$ and $\omega = 14.3$ in the semi-infinite band gap, due to an applied moment $\vec{f} = [0,-1]^{\mathrm{T}}$ at $n=0$.}
     \label{1DseminfVDRF}
\end{minipage}
\end{figure}

Alongside studying the finite band gap, we also construct Green's functions for the semi-infinite band gap. 
Using the same value of the rotational inertia, the band gap Green's function for the flexural displacement in the case of translational forcing $\vec{f} = [-1,0]^{\mathrm{T}}$ has been plotted in Figure~\ref{1DseminfVDVF} for $\omega=14.3$.
Similarly, we evaluate the Green's function for the flexural displacement due to an applied moment $\vec{f} = [0,-1]^{\mathrm{T}}$. 
This has been plotted in Figure~\ref{1DseminfVDRF} and exhibits the signature twist shape as expected.

Comparing Figures \ref{vertmotionplot} and \ref{1DseminfVDVF}, 
it is seen that the displacements have significantly different shape despite having the same type of forcing.
The same can be said comparing Figures \ref{EBCfiniteW9,8Rot} and \ref{1DseminfVDRF}.
These differences can be explained by looking a the dispersion curves in Figure~\ref{figbeamchaindisp}. 
For values of $\omega$ in the finite band gap, the complex solutions of the dispersion equation are of the form $k = \pi + \I \hat{k}$, where $\hat{k}\in\mathbb{R}$, leading to oscillatory eigenmodes of the form $e^{i\pi n - \hat{k}n}$.
In contrast, for the semi-infinite band gap, the complex solutions of the dispersion equation are purely imaginary, leading to eigenmodes of the form $e^{-\hat{k} n}$ associated with non-oscillatory evanescent waves.


\subsection{Pass band frequencies}

\begin{figure}[ht]
\centering
\begin{minipage}{\textwidth}
    \centering
    \includegraphics[width=0.85\textwidth]{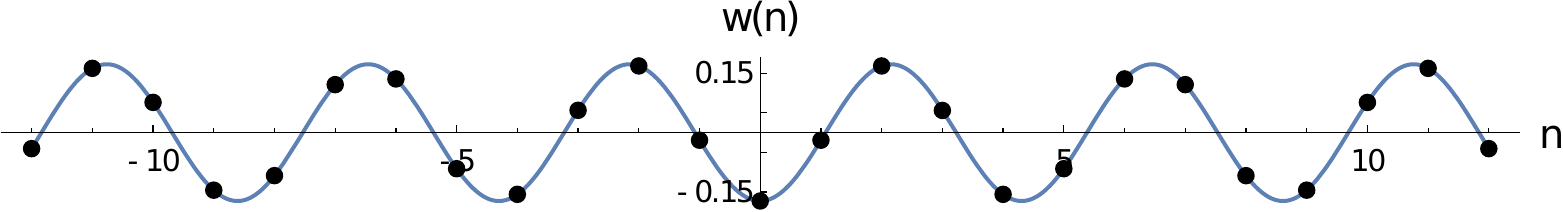}
    \caption{A plot of the flexural displacement Green's function $w(n)$ for $\mu=0.06$ and $\omega = 2$ in the lower pass band, due to point forcing $\vec{f} = [-1,0]^{\mathrm{T}}$ at $n=0$.}
     \label{fig1DchainPASSw(x)VertForce}
\end{minipage}
\end{figure}

While the Euler--Bernoulli beam chain has two pass bands, we note that in both the upper and lower pass bands, the position of the poles in relation to the unit circle, from equation~\eqref{eq1DResidueeq}, remains unchanged for fixed $\mu$. 
Thus we can form one Green's function for both of the pass band regimes, which remains frequency dependent. 
The flexural displacement Green's function in response to a translational point force at $n=0$ of frequency $\omega=2$, is plotted in Figure~\ref{fig1DchainPASSw(x)VertForce}.
As expected, we note that waves of constant amplitude are observed, indicating propagative modes.

\subsection{Rotational displacement}

Evaluating the inverse Fourier transformation of $\Theta^{F}(k)$ component of $\vec{u}^{F}(k)$ in equation~\eqref{eqmatrixoperator} gives the rotational displacement of the chain $\theta_{x}$ in response to the applied force.
Until now, we have emphasised the flexural displacement component since the plots of $w(n)$ can be interpreted intuitively as the deformed shape of the chain.
However the rotational displacement Green's function is a useful tool that provides further insight into the lattice behaviour, and a quantitative measure of how much each mass rotates.

In Figure~\ref{EBCfiniteROTDISw9,8Vert} we plot $\theta_{x}(n)$ in response to a translational point force $\vec{f} = [-1,0]^{\mathrm{T}}$ for $\omega = 7.5$ and $\mu = 0.06$. These are the same conditions applied that were applied to Figure~\ref{vertmotionplot} when plotting the flexural displacement. 
In Figure~\ref{EBCfiniteROTDISw9,8Vert}, it can be seen that the $n=0$ mass has zero rotational displacement under purely translational forcing, which is expected at the central node of the lattice where the force is applied. 
This agrees with Figure~\ref{vertmotionplot} which shows that while the beams either side of the $n=0$ mass deform, the mass itself does not twist in place. 
It is also seen that the magnitude of the rotational displacement of each node decreases as $n$ increases, which is characteristic of a decaying wave.
\begin{figure}[ht]
\centering
\begin{minipage}{\textwidth}
    \centering
    \includegraphics[width=0.85\textwidth]{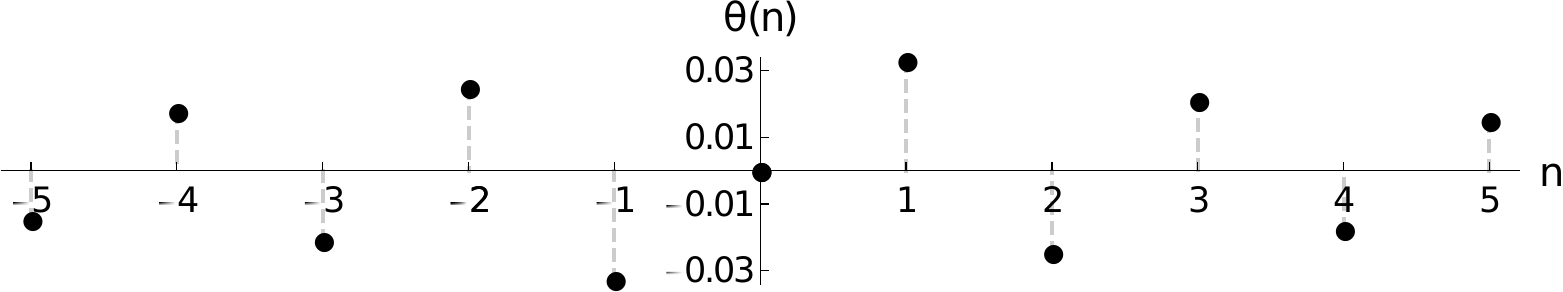}
    \caption{A plot of the rotational displacement $\theta_{x}(n)$ of each node, for the values of $\mu=0.06$ and $\omega = 7.5$ within the finite band gap, due to point forcing $\vec{f} = [-1,0]^{\mathrm{T}}$ at $n=0$. }
     \label{EBCfiniteROTDISw9,8Vert}
\end{minipage}
\end{figure}


\subsection{Degenerate point}\label{SECDegeneratemu}

As shown in \S~\ref{secdispeq}, the finite band gap collapses for only one value of the rotational inertia, $\mu = 1/12$.
When $\mu = 1/12$, the dispersion equation degenerates and the two simple poles associated with the Green's function coalesce at $\omega=4\sqrt{3}$. 
At this point of degeneracy, for a translational point force at $n=0$, the inverse Fourier transformation of the flexural displacement has a convenient representation in terms of the PFQ regularised hypergeometric function $\tilde{F}$ as follows
\begin{equation}
    w(n) = \frac{1}{2\pi}\int\limits_{-\pi}^{\pi} \frac{\mathrm{e}^{\I kn}}{12\, (\cos(k)-5)} \dd k  = -\frac{1}{36} \pi  \,  _3\tilde{F}_2  \left(\frac{1}{2},1,1;1-n,n+1;\frac{1}{3}\right).
\end{equation}
The displacement $w(n)$ at the degeneracy results in an oscillatory evanescent wave, despite the frequency $\omega=4\sqrt{3}$ being a solution to the dispersion equation, that is $\sigma(4\sqrt{3},\pi)=0$, which would indicate a propagative mode. 
Other studies with detail on evaluating Green's functions using hypergeometric functions are found in~\cite{lattice-sums,ColquittLineOfDefects} and references therein.


\section{General Method}\label{setup}
In this section we extend the previous work to form a general method for the construction of dynamic Green's functions on $d$-dimensional lattices with vector connections capable of describing a wide range of mechanical interactions, including elastic beams, rods, and springs. 
This method allows for the inclusion of interesting effects, such as the torsional stiffness and out-of-plane flexural deformations of the lattice connections, as was demonstrated in earlier sections.

The equations of motion are of the form 
\begin{equation}
    \mathsf{D} \vec{u}_\vec{n}(t) + \mathsf{M}\ddot{\vec{u}}_\vec{n} (t) = \vec{f}_{\vec{n}},
\end{equation}
where $\vec{n}\in\mathbb{Z}^d$ enumerates the lattice points in which the inertia of the system is concentrated, $d\in\mathbb{N}\setminus\{0\}$ is the dimension of space, $\vec{u}_\vec{n}\in\mathbb{C}^d$ denotes the generalised displacement which includes translation and rotation of the $\vec{n}^{th}$ node, $\mathsf{D}$ is a discrete [tensor] operator which encapsulates the interaction (e.g. compressional, torsional, flexural, stiffness) of the lattice connections and $\mathsf{M}$ is a tensor describing the inertial properties of the system. 
The vector $\vec{f}_{\vec{n}} \in \mathbb{C}^d$ describes the [generalised] forcing --- which incorporates, for example, linear forces, bending moments, and torsional moments --- and $\ddot{\vec{u}}_\vec{n}$ is the second derivative of $\vec{u}_{\vec{n}}$ with respect to time.
In the absence of forcing and for time-harmonic motion, of angular frequency $\omega$, the equations of motion reduce to
\begin{equation}
    \mathsf{D} \vec{u}_\vec{n} -\omega^2 \mathsf{M}\vec{u}_\vec{n} = \vec{0},
\label{eqgeneqofmotion}
\end{equation}
where $\vec{0}$ is the $d$-dimensional zero vector, and the common factor of $\mathrm{e}^{\I\omega t}$ has been omitted for convenience.
We introduce the discrete Fourier transformation,
\begin{equation}
\vec{u}^{F}(\vec{k}) = \sum_{\vec{n}\in\mathbb{Z}^d} \mathrm{e}^{-\I \vec{k}\cdot\vec{n}} \vec{u}_\vec{n} \, ,
\label{eqgeneralFT}
\end{equation}
where $\vec{k}$ is the $d$--dimensional Fourier variable.
The equation of motion in reciprocal space is provided by the Fourier transformation of equation~\eqref{eqgeneqofmotion}, 
\begin{equation}
    [\mathsf{L}(\vec{k}) - \omega^{2}\mathsf{M}]\vec{u}^{F}(\vec{k}) = \vec{0},
    \label{eqGeneralOP}
\end{equation}
where we define $\mathsf{L}(\vec{k})$ as the Fourier transformed operator $\mathsf{D}$.
The solvability condition of equation~\eqref{eqGeneralOP} leads to the dispersion equation
\begin{equation}
    \det\left[ \mathsf{L}(\vec{k}) - \omega^{2}\mathsf{M}\right] = 0.
\end{equation}
For discrete systems this equation is polynomial in $\omega$ and the solutions of the dispersion equations can be plotted to produce dispersion diagrams 
and identify the regions where no real solutions to the dispersion equation exist, creating band gaps.
In general, exciting the system at a frequency within a band gap will yield localised modes that decay exponentially with distance from the forcing point.

We construct the Green's functions by considering the action of time-harmonic generalised forces at the origin of the $d$-dimensional lattices.
In this case, the equations of motion are
\begin{equation}
    \mathsf{D}\vec{u}_\vec{n} - \omega^{2} \mathsf{M} \vec{u}_\vec{n} = \delta_{\vec{n},\vec{0}}\vec{f_{n}} \,,
    \label{eq:gf-eom}
\end{equation}
where $\delta_{\vec{n},\vec{m}}$ is the Kr\"onecker delta, and we denote $\vec{f_{0}}=\vec{f}$.
Following the application of the Fourier transformation, equation~\eqref{eq:gf-eom} becomes
\begin{equation}
    [\mathsf{L}(\vec{k}) - \omega^{2}\mathsf{M}]\vec{u}^{F}(\vec{k}) = \vec{f}. \label{eqGeneralGreenFT}
\end{equation}
The Green's functions describing the real space displacements $\vec{u}_\vec{n}$ are determined by applying the inverse of the discrete Fourier transformation as follows, 
\begin{equation}
    \vec{u}_{\vec{n}} = \frac{1}{(2\pi)^{d}}\int\limits_{-\pi}^{\pi} \mathrm{e}^{\I \vec{k} \cdot \vec{n}}\; \vec{u}^{F}(\vec{k})\,  \dd \vec{k} \,=\, \frac{1}{(2\pi)^{d}}\int\limits_{-\pi}^{\pi}  \mathrm{e}^{\I \vec{k} \cdot \vec{n}} \left[\mathsf{L}(\vec{k}) - \omega^{2}\mathsf{M} \right]^{-1} \vec{f}\, \dd \vec{k}\, .
    \label{eqgenintegral}
\end{equation}

Now, rather than applying external forces to the lattice systems, we consider creating localised modes through the introduction of an isolated defect in the inertial properties of the lattices at $\vec{n} = \vec{0}$. In each system the defect is denoted by the $d$-dimensional tensor $\mathcal{M}$ which describes the altered inertial properties at the origin.
The Fourier transformed equations of motion for lattices with a defect at the origin are then
\begin{equation}
    [\mathsf{L}(\vec{k}) - \omega^{2}\mathsf{M}]\vec{u}^{F}(\vec{k}) = \mathcal{M} \omega^{2}\vec{u}_{\vec{0}}. \label{eqMassdefectEoM}
\end{equation}
By choosing an $\mathcal{M}$ such that 
\begin{equation}
    \mathcal{M} \omega^{2}\vec{u_{0}} = \vec{f},
    \label{eqmassdefect}
\end{equation}
it is possible to recover equation~\eqref{eqGeneralGreenFT} and, in doing so, one can produce localised eigenmodes that coincide with band gap Green's functions. 
The displacement field corresponding to the localised defect modes can be found from equation~\eqref{eqMassdefectEoM}, using the inverse Fourier transformation.


\subsection{A mass defect for the Euler-Bernoulli beam chain}\label{sec1Ddefectmode}

Using the above method, we derive a localised defect $\mathcal{M}$ for the Euler--Bernoulli beam chain such that the chain supports a localised mode for a chosen band gap frequency $\omega$. 
The Green's functions $w(n)$ and $\theta_{x}(n)$, which form the vector $\vec{u}_{n}$ can be evaluated in closed form for band gap frequencies using equation~\eqref{eq1DResidueeq} and the analogous equation for $\theta_{x}$.
Thus, $\vec{u}_{0} = [w(0), \theta_{x}(0)]^{\mathrm{T}}$ from equation~\eqref{eqmassdefect} is in-hand. 

It is noticed that in the case of a purely translational force, the central node experiences zero rotational displacement, that is, $\theta_{x}(0) = 0$ when $\vec{f}=[-1,0]^{\mathrm{T}}$, as can be seen explicitly in Figure~\ref{EBCfiniteROTDISw9,8Vert}. 
Likewise in the case of an applied moment $\vec{f}=[0,-1]^{\mathrm{T}}$, we see that $w(0)=0$, which is demonstrated in Figure~\ref{EBCfiniteW9,8Rot}.
To avoid singular matrices, we define the functions $\alpha_{1}$ and $\alpha_{2}$ as 
\[
    \alpha_{1} = 
    \begin{cases}
    1 & \text{if } f_{w} = 0\\[1em]
    \dfrac{f_{w}}{\omega^{2}w(0)} & \text{if } f_{w} \neq 0
    \end{cases}
    \qquad\text{and}\qquad
    \alpha_{2} = 
    \begin{cases}
    \mu & \text{if } f_{\theta} = 0\\[1em]
    \dfrac{f_{\theta}}{\omega^{2}\theta_{x}(0)} & \text{if } f_{\theta} \neq 0 \,. 
    \end{cases}
\]
As such, we define the inertial defect at the central node $n=0$ using the matrix
\begin{equation}
\mathcal{M} = \begin{bmatrix}
\alpha_{1}  &  0\\
  0   &  \alpha_{2}
  \end{bmatrix},
\label{1DMassDefectVerticalF}
\end{equation}
which is dependent on the applied forcing, and will result in a localised mode around the origin for the chosen band gap frequency.
The function $\alpha_{1}$ corresponds to a change in the mass of the central node $n=0$, while $\alpha_{2}$ corresponds to the changing the rotational inertia.  
Off diagonal elements in the matrices would correspond to coupling between the mass and rotational inertia; coupling of parameters is studied in other works such as~\cite{CoupledStressCloak} but is not relevant for this problem. 


\subsection{A mass defect for the square Euler-Bernoulli beam lattice}\label{Sec2DmassDefect}

In the same manner as above, we can evaluate the inertial defect required for the 2D square lattice to support a localised mode for a chosen band gap frequency $\omega$, coincident with the band gap Green's function for the desired forcing. 
Since the Green's function $\vec{u}_{{(m,n)}}$ can be evaluated numerically at the $\vec{0}$ node, it remains possible for us to find the required values of $w(0,0)$, $\theta_{x}(0,0)$ and $\theta_{y}(0,0)$ that compose the $\vec{u}_{(0,0)}$ vector. 
As above, to avoid singular matrices, we define the functions
\[
    \beta_{1} = 
    \begin{cases}
    1 & \text{if } f_{w} = 0\\[1em]
    \dfrac{f_{w}}{\omega^{2}w(0,0)} & \text{if } f_{w} \neq 0,
    \end{cases}
    \qquad\qquad
    \beta_{2} = 
    \begin{cases}
    \mu & \text{if } f_{\theta_{x}} = 0\\[1em]
    \dfrac{f_{\theta_{x}}}{\omega^{2}\theta_{x}(0,0)} & \text{if } f_{\theta_{x}} \neq 0. 
    \end{cases}
\]
\[
    \text{and} \quad
    \beta_{3} = 
    \begin{cases}
    \mu & \text{if } f_{\theta_{y}} = 0\\[1em]
    \dfrac{f_{\theta_{y}}}{\omega^{2}\theta_{y}(0,0)} & \text{if } f_{\theta_{y}} \neq 0. 
    \end{cases}
\]
Following this, we define the inertial defect matrix as
\begin{equation}
    \mathcal{M} = 
    \begin{bmatrix}
    \beta_{1} & 0 & 0\\
    0 & \beta_{2} & 0\\
    0 & 0 & \beta_{3}
    \end{bmatrix} ;
    \label{2dinverseMdefectV}
\end{equation}
resulting in a localised defect mode about the $\vec{0}$ node. 
The component $\beta_{1}$ corresponds to a defect in the mass of the $\vec{0}$ node, while the $\beta_{2}$ and $\beta_{3}$ components correspond to defects in the rotational inertia for the $x$- and $y$-directions respectively.


\section{Concluding Remarks}

Firstly we studied a 2D square lattice of Euler--Bernoulli beams with junctions that possessed both mass and rotational inertia. The beams also possessed torsional stiffness, to account for the coupling between flexural and torsional waves at the junctions. 
The Green's function was used alongside finite element software to illustrate the extreme anisotropy (including remarkable asymmetric anisotropy, usually associated with PT-symmetry breaking) that can be achieved, and controlled in the lattice by altering the frequency and nature of the applied forcing. 
Analysis of the dispersion diagrams for various values of the rotational inertia and torsional stiffness was used to illustrate how the propagation of waves and dynamic behaviour of the lattice can be manipulated, giving an unprecedented level of control over the pass band frequencies and their preferential directions through the lattice. 

We also studied the related problem of a 1D chain of Euler--Bernoulli beams with junctions that possess mass and rotational inertia. 
Closed form analytical Green's functions were achieved for each of the pass band and band gap regimes of the chain, for different choices of forcing. 
We also demonstrated how the rotational inertia provides significant control over the propagating frequencies on the lattice. 
Lastly, we provided a method to construct Green's functions for a generalised $d$-dimensional lattice with discrete mass and arbitrary linear interactions between nodes.
We then provided a method to design lattice defects such that the lattice possesses a localised mode of a chosen frequency.
This method was used to derive inertial defect matrices for the 2D lattice and the 1D chain from earlier in the paper.  
The work in this paper has applications in many areas where the control of wave propagation through elastic structures is of particular interest --- including metamaterials, seismic protection, energy dissipation, and others.

\section*{Acknowledgements}
Financial support from the University of Liverpool and National Tsing Hua University, through its dual-PhD programme, to KM is gratefully acknowledged.
The authors are also grateful for valuable conversations with Professor Natasha Movchan regarding the structure of the paper.

{\small \bibliography{1Abibliography}}

\end{document}